\documentclass[prl,aps,onecolumn,superscriptaddress,notitlepage]{revtex4}
\usepackage{graphicx}
\usepackage{amsmath}
\usepackage{times}
\usepackage{color}
\usepackage{multirow}
\usepackage{subfigure}

{\left\lbrace\begin{array}{@{}l@{}}}%
{\end{array}\right.} 

\newcommand{\ket}[1]{|#1\rangle}

\begin{document}

\title{Three-photon bosonic coalescence in an integrated tritter}

\author{Nicol\`o Spagnolo}
\affiliation{Dipartimento di Fisica, Sapienza Universit\`{a} di Roma, Piazzale Aldo Moro 5, I-00185 Roma, Italy}
\author{Chiara Vitelli}
\affiliation{Center of Life NanoScience @ La Sapienza, Istituto Italiano di Tecnologia, Viale Regina Elena, 255, I-00185 Roma, Italy}
\affiliation{Dipartimento di Fisica, Sapienza Universit\`{a} di Roma, Piazzale Aldo Moro 5, I-00185 Roma, Italy}
\author{Lorenzo Aparo}
\affiliation{Dipartimento di Fisica, Sapienza Universit\`{a} di Roma, Piazzale Aldo Moro 5, I-00185 Roma, Italy}
\author{Paolo Mataloni}
\affiliation{Dipartimento di Fisica, Sapienza Universit\`{a} di Roma, Piazzale Aldo Moro 5, I-00185 Roma, Italy}
\author{Fabio Sciarrino}
\email{fabio.sciarrino@uniroma1.it}
\affiliation{Dipartimento di Fisica, Sapienza Universit\`{a} di Roma, Piazzale Aldo Moro 5, I-00185 Roma, Italy}
\author{Andrea Crespi}
\affiliation{Istituto di Fotonica e Nanotecnologie, Consiglio Nazionale delle Ricerche (IFN-CNR), Piazza Leonardo da Vinci, 32, I-20133 Milano, Italy}
\affiliation{Dipartimento di Fisica, Politecnico di Milano, Piazza Leonardo da Vinci, 32, I-20133 Milano, Italy}
\author{Roberta Ramponi}
\affiliation{Istituto di Fotonica e Nanotecnologie, Consiglio Nazionale delle Ricerche (IFN-CNR), Piazza Leonardo da Vinci, 32, I-20133 Milano, Italy}
\affiliation{Dipartimento di Fisica, Politecnico di Milano, Piazza Leonardo da Vinci, 32, I-20133 Milano, Italy}
\author{Roberto Osellame}
\email{roberto.osellame@polimi.it}
\affiliation{Istituto di Fotonica e Nanotecnologie, Consiglio Nazionale delle Ricerche (IFN-CNR), Piazza Leonardo da Vinci, 32, I-20133 Milano, Italy}
\affiliation{Dipartimento di Fisica, Politecnico di Milano, Piazza Leonardo da Vinci, 32, I-20133 Milano, Italy}

\maketitle

\textbf{The main features of quantum mechanics reside in interference deriving from the superposition of different quantum objects. While current quantum optical technology enables two-photon interference both in bulk and integrated systems, simultaneous interference of more than two particles, leading to richer quantum phenomena, is still a challenging task. Here we report the experimental observation of three-photon interference in an integrated three-port directional coupler realized by ultrafast laser-writing. By exploiting the capability of this technique to produce three-dimensional structures, we realized and tested in the quantum regime a three-port beam splitter, namely a tritter, which allowed us to observe bosonic coalescence of three photons. These results open new important perspectives in many areas of quantum information, such as fundamental tests of quantum mechanics with increasing number of photons, quantum state engineering, quantum sensing and quantum simulation.}

\section{Introduction}

The Hong-Ou-Mandel effect is a purely quantum phenomenon deriving from the interference of two photons \cite{Hong87}. It is a consequence of the bosonic nature of light and it can be observed when two identical photons impinge simultaneously on a two-port balanced beam splitter and, as a result, they are forced to exit from the same port. This effect can be exploited to quantify the amount of indistinguishability of correlated photons \cite{Ou06, Yama03} or to project a quantum state onto the maximally-entangled singlet Bell state, representing an important resource in many research areas of quantum information, from quantum computation to quantum metrology \cite{Walt04, Kok07, Knil00}. 
The interaction of single-photon states on a beam splitter and successive post-selection may indeed lead to the generation of entangled states \cite{Shih88} such as N00N and GHZ states \cite{XuBo02}. As a matter of fact, achieving interference of more than two particles on a beam splitter is a challenging task \cite{Gree93}, however it could open the door to the observation of quantum phenomena in increasing size systems and to the engineering of quantum states for quantum information processing \cite{Gree00,Pan12}.

%
\begin{figure}[ht!]
\centering
\includegraphics[width=0.5\textwidth]{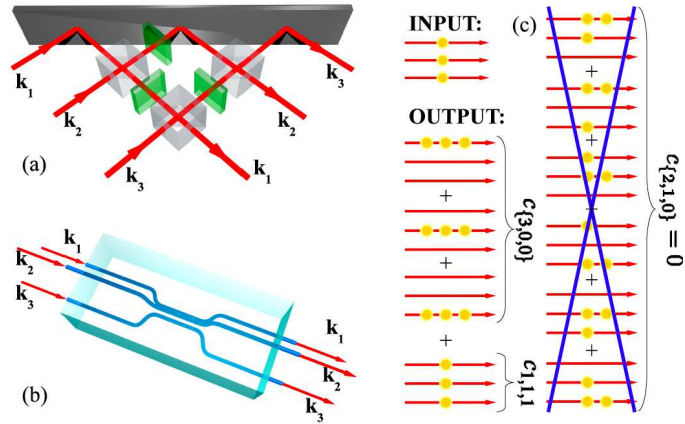}
\caption{{\bf Schematic representation of a tritter and related bosonic coalescence.} (a) Optical scheme for a three-mode splitter obtained by a set of conventional beam-splitters (grey cubes) and phase shifters (green plates) \cite{Zuko97,Camp00}, where $\mathbf{k_{i}}$ with $i=1,2,3$ represent input-output spatial modes. (b) Integrated 3-dimensional scheme for a laser-written tritter. (c) Three-photon bosonic coalescence effect in a tritter injected by one photon over each different input spatial modes: all the terms with two photons exiting on the same output port vanish due to bosonic coalescence.}
\label{fig:chip}
\end{figure}
%

The adoption of a multiport beam splitter has been proposed to perform additional classes of possible EPR-Bell experiments in a Hilbert space with dimension higher than two \cite{Zuko97}. Furthermore the capability of implementing multipath interferometers by multiport beam splitters could be adopted to perform quantum computation tasks \cite{Howe00}. Recently a generalized $N$-port beam splitter has been theoretically addressed in Ref. \cite{Reck94,Tich10}, and the behaviour of $N$ photons impinging onto the $N$-port beam splitter has been studied by using a statistical approach, since the complexity of this problem grows faster than exponentially with growing $N$. Very recently, the application to quantum interferometry of three- and four-port beam splitters have been theoretically investigated, and the quantum enhancement in terms of sensitivity has been predicted \cite{Spag12}. However, three particle interferometry is mostly unexplored territory, both experimentally and theoretically. The three beam extension of a beam splitter, the so called `tritter', was first addressed in Ref. \cite{Zuko97} and its proposed implementation consisted in the combination of two-port beam splitters and phase shifters [see Fig. \ref{fig:chip} (a)] \cite{Zuko97,Camp00}. Three dimensional implementations of  three- and four-port beam splitters, based on fused optical fibers \cite{Weih96} or femtosecond laser waveguide writing \cite{Kowa05,Suzu06,With12}, have been proposed. However, only classical light or two-photon interference experiments were carried out on such devices. Therefore, an experimental investigation of quantum interference of more than two-photons is still lacking. 

Here we report on the first experimental investigation of three-photon interference in a tritter. The tritter is realized by femtosecond laser waveguide writing and it consists of three waveguides approaching three-dimensionally, thus enabling simultaneous interaction of the three photons without having to decompose the process into cascaded two-mode interactions and phase shifters [Fig. \ref{fig:chip} (b)]. The three photons, each coupled to a different waveguide, mutually interfere in the three-arm directional coupler by evanescent field interaction. The behaviour of a three-photon state entering the tritter is investigated by comparing the output state probabilities with the expected classical ones, evidencing the purely quantum effects involved in bosonic coalescence.

%
\begin{figure}[ht!]
\centering
\includegraphics[width=0.99\textwidth]{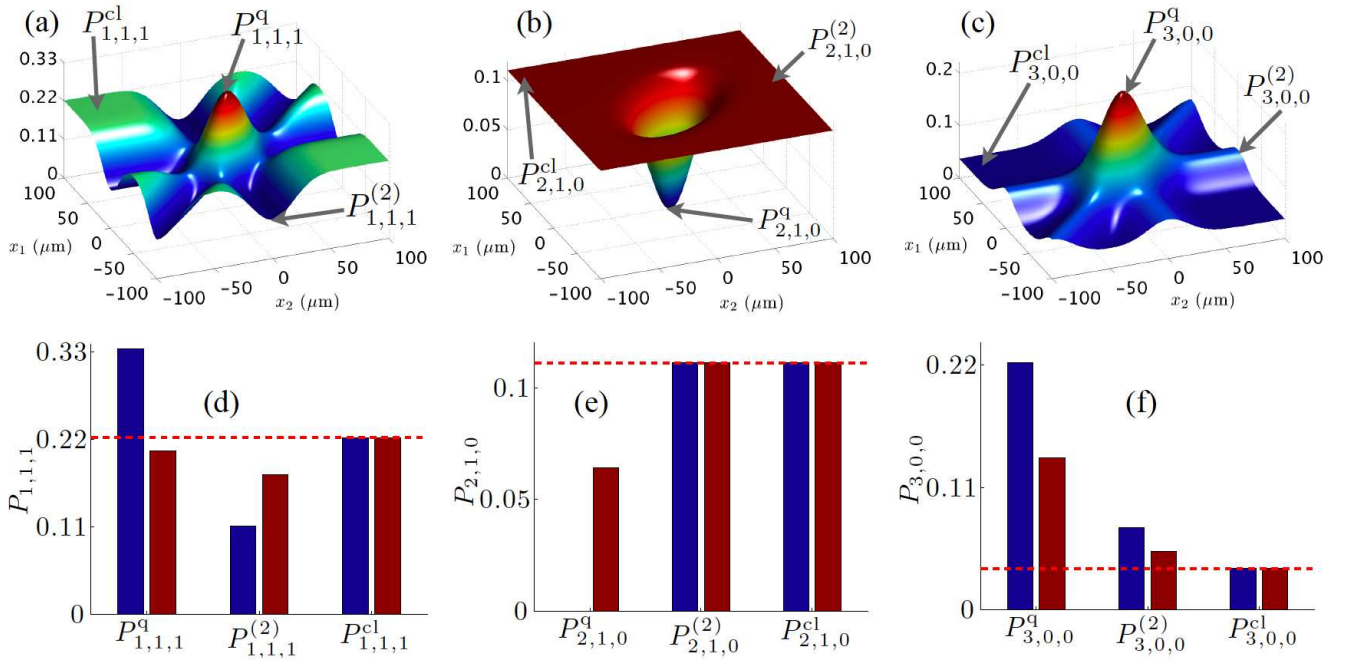}
\caption{{\bf Hong-Ou-Mandel effect for a $\vert 1,1,1 \rangle$ input state in an ideal tritter}. (a-c) Theoretical output probabilities $P_{1,1,1}$ (a), $P_{2,1,0}$ (b), $P_{3,0,0}$ (c) as a function of the delays $x_{i}=c \tau_{i}$ of the impinging photons. (d-f) Theoretical output probabilities $P^{\mathrm{q}}_{i,j,k}$ for three indistinguishable input photons, $P^{\mathrm{cl}}_{i,j,k}$ for three photons out of interference, and $P^{(2)}_{i,j,k}$ for two indistinguishable and one delayed input photons. The red bars take into account the presence of partial distinguishability between the three impinging photons with indistinguishability $p=0.65$ while blue bars refer to perfectly indistinguishable photons with $p=1$. (see Supplementary Note 4A). The dashed horizontal lines represent the classical boundaries.}
\label{fig:oumandel}
\end{figure}
%

\section{Results}

\textsl{\textbf{Theoretical description of the tritter output probabilities}}. - Let us first address the action of a tritter on a generic input state $\vert \psi \rangle$. It can be expressed by a unitary matrix $\mathcal{U}$, mapping the input field operators $a^{\dag}_{i}$ to the output field operators $b^{\dag}_{i}$ with $b^{\dag}_{i} = \sum_{j} \mathcal{U}_{ij} a^{\dag}_{j}$, \cite{Zuko97,Camp00,Tich10} (see Supplementary Note 1A). In the case of an ideally symmetric tritter, where a photon entering in one input port has the same probability of exiting through anyone of the output ports, the unitary matrix $\mathcal{U}$ reads: 
\begin{equation}
\label{eq:tritter_ideal}
\mathcal{U}^{t} = \frac{1}{\sqrt{3}} \begin{pmatrix} 1 & 1 & 1\\ 1 & e^{\imath 2 \pi/3} & e^{\imath 4 \pi/3}\\ 1 & e^{\imath 4 \pi/3} & e^{\imath 8 \pi/3} \end{pmatrix}.
\end{equation}
Let us consider what happens when an ideal tritter is injected by a three-photon state $\vert 1,1,1\rangle$. The corresponding output state after the evolution induced by $\mathcal{U}^{t}$ reads:
\begin{equation}
\vert 1,1,1 \rangle \stackrel{\mathcal{U}^{t}}{\rightarrow} c_{1,1,1} \vert 1,1,1 \rangle + c_{\{3,0,0\}} \vert \{3,0,0\} \rangle.
\end{equation}
Here $c_{1,1,1}=-1/\sqrt{3}$ and $c_{\{3,0,0\}}=\sqrt{2/3}$, where $\vert \{i,j,k\} \rangle$ is the symmetric superposition of the three-photon states corresponding to $(i,j,k)$ photons exiting in the three output ports. We first observe that, as a consequence of bosonic coalescence, all the terms of two photons exiting from the same output port disappear [Fig. \ref{fig:chip} (c)]. In order to fully describe three-photon interference we need to introduce two relative delays between the three bosons. We theoretically calculated the probability of the different output states [$(1,1,1)$, $(2,1,0)$, $(3,0,0)$] as a function of the relative delays $x_{m} = c \tau_{m}$ $(m=1,2)$ between the three photons. These probabilities are represented in Figs. \ref{fig:oumandel} (a-c) as surfaces in a three-dimensional space and represent an extension of the well-known Hong-Ou-Mandel curves. This representation enables a deeper insight into three-photon interference, evidencing non-trivial features as a function of the two delays. In particular, we observe the presence of three regions: (i) the three input photons are indistinguishable $(P_{i,j,k}^{\mathrm{q}})$ when $x_{1} \sim x_{2} \sim 0$, this leads to a three-photon coalescence effect, resulting in a dip or a peak depending on the detected output state; (ii) only two of the three input photons are indistinguishable $(P_{i,j,k}^{(2)})$ when $x_{m} \sim 0 \neq x_{n}$ or $x_m \sim x_n$; (iii) the three input photons are distinguishable $(P_{i,j,k}^{\mathrm{cl}})$ when $x_{1}\neq x_{2} \neq 0$, leading to classically correlated outcomes. Here, the superscript $q$ indicates the quantum nature of the three-photon coalescence, while the superscript $cl$ highlights that the photons behave as classical particles. Figs. \ref{fig:oumandel} (d-f) compare the different probabilities corresponding to cases (i-iii) for each output state.

%
\begin{figure*}[ht!]
\centering
\includegraphics[width=0.99\textwidth]{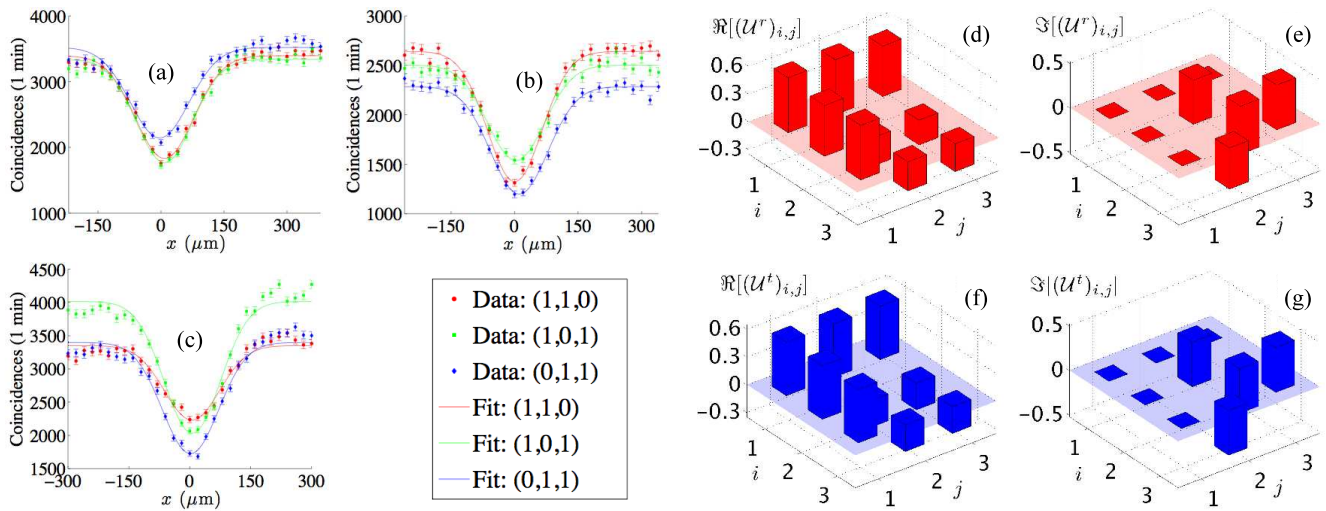}
\caption{{\bf Two-photon experimental reconstruction of the unitary matrix $\mathcal{U}^{r}$ of the implemented tritter.} (a)-(c) Experimental two-photon Hong-Ou-Mandel interference for input states: (a) $\vert 1, 1, 0\rangle$, (b) $\vert 1, 0, 1\rangle$, (c) $\vert 0, 1, 1\rangle$. For each input state all possible output ones are acquired.  Error bars in the experimental points are due to poissonian statistics of the measured signal. (d)-(e) Real and imaginary parts of $\mathcal{U}^{r}$. (f)-(g) Real and imaginary parts of the theoretical tritter $\mathcal{U}^{t}$.}
\label{fig:reconstruction}
\end{figure*}
%

\textsl{\textbf{Tritter fabrication by ultrafast laser writing}}. - The tritter device was realized by adopting femtosecond laser writing technology \cite{Gatt08, Dell09,Osel11}. This technique exploits nonlinear absorption of focused femtosecond laser pulses that create permanent and localized increase of the refractive index in transparent materials. Waveguides are directly fabricated in the material bulk by translating the sample at constant speed along the desired path. This technique has the unique capability of producing integrated devices with three-dimensional geometries. Ultrafast laser-writing technique has rapidly become a powerful tool for demonstrating new quantum integrated devices, able to perform quantum logic operations \cite{Cres11} as well as two-photon quantum walks \cite{Owen11, Sans12}. The femtosecond laser written tritter consists of three waveguides approaching and interacting in the coupling region [Fig. \ref{fig:chip} (b)]. Thanks to this configuration equal coupling coefficients may be obtained, in a way that a single photon entering in any input port has the same probability to exit from any of the three output ports. This three-dimensional approach, enabled by femtosecond laser writing, avoids the use of cascaded beam splitters and phases which are difficult to control and provides a compact device that can be effectively scaled in more complex architectures. Although conceptually simple, the tritter needs accurate dimensioning. As a first order arrangement we consider the waveguides lying at the vertex of an equilateral triangle, where key parameters are waveguide spacing and length in the interaction region. Actually these two parameters are not independent since, in order to achieve and equal output distribution, the coupling coefficient $k$ and interaction length $L$ need to satisfy the relation $k L=2\pi/9$ (see Supplementary Note 2). Sizing of the tritter starts from choosing the waveguide spacing $d$ to the smallest value that avoids overlap (this will keep the device as short as possible). The chosen value of $d$ in turn defines the coupling coefficient $k$ and thus, according to the previous relation, the interaction length $L$. However, the slightly elliptical guided mode of the femtosecond laser written optical waveguides and other fabrication imperfections require performing a second order fine tuning of the device that can be accomplished by slightly shifting one of the three waveguides in a scalene triangle geometry. This fine tuning is performed most efficiently by an iterative procedure where we take full advantage of the rapid prototyping capabilities of femtosecond laser waveguide writing (see Supplementary Figure S1 for the final optimized geometry).

\textsl{\textbf{Two-photon characterization of the tritter device}}. - To characterize the implemented device and to compare its unitary evolution with respect to the ideal case, we implemented the reconstruction method proposed in Ref. \cite{Peru11}. This technique consists in measuring all the possible two-photon Hong-Ou-Mandel interference fringes and then reconstructing the elements of the evolution matrix $\mathcal{U}^{r}$ through a minimization procedure (see Supplementary Note 3 for more details). We show in Fig. \ref{fig:reconstruction} the real and imaginary parts of the reconstructed matrix $\mathcal{U}^{r}$ at 795 nm wavelength, compared to the corresponding theoretical matrix $\mathcal{U}^{t}$. To quantify the quality of the fabrication process we evaluated the similarity $S$ between the two corresponding Hong-Ou-Mandel visibility matrices, defined as $\mathcal{S}_{r,t}=1-\sum_{i\neq j} \sum_{k \neq l} \vert (\mathcal{V}^{t})_{i,j;k,l} - (\mathcal{V}^{r})_{i,j;k,l} \vert/18$. The obtained value $\mathcal{S}_{r,t}=0.9768 \pm 0.0004$ demonstrates a high correspondence of the implemented device with an ideal tritter.

%
\begin{figure*}[ht!]
\centering
\includegraphics[width=0.99\textwidth]{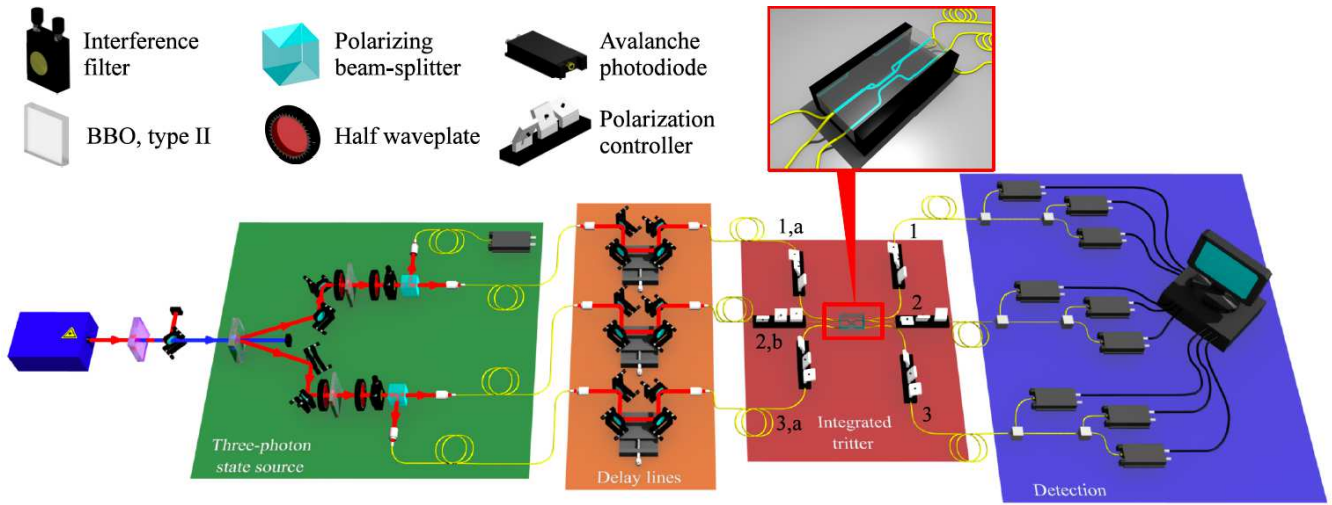}
\caption{{\bf Experimental layout}. Experimental setup for the characterization of the tritter and the three-photon coalescence experiment. Two-photon and three-photon states, generated by parametric down-conversion, are injected in the tritter device after the propagation through spatial delay lines. Then, coincidence detection at the output ports are performed to reconstruct the probability of obtaining a given output state realization $(i,j,k)$. Labels $1,2,3$ stand for spatial modes, while letter $a$ refers to indistinguishable photons and $b$ to the partially distinguishable one.}
\label{fig:chip_exp}
\end{figure*}
%

\begin{figure}[ht!]
\centering
\includegraphics[width=0.99\textwidth]{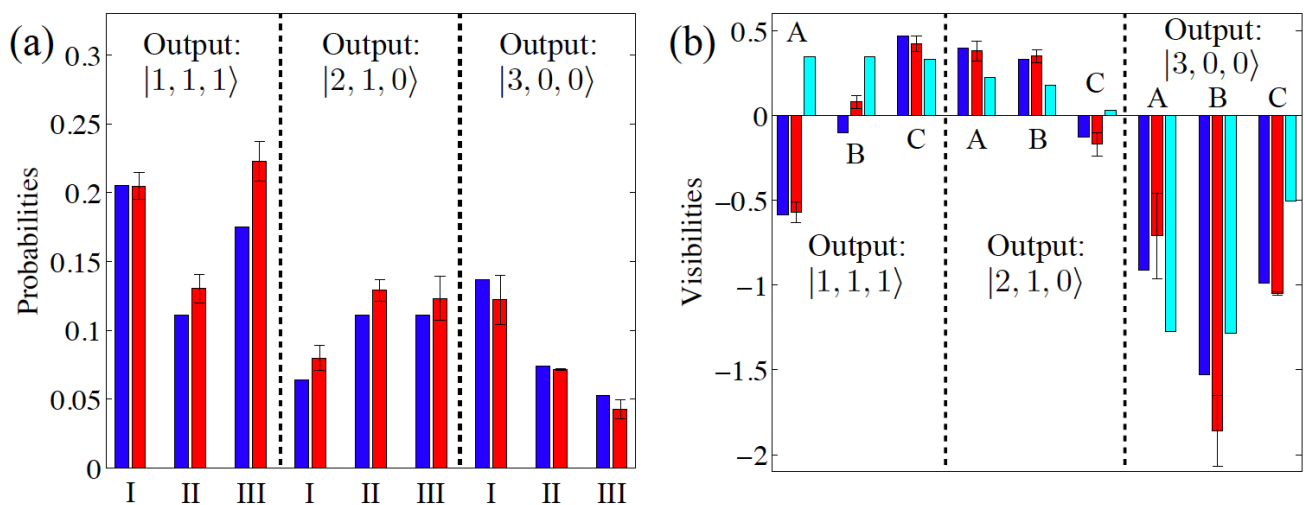}
\caption{{\bf Experimental results with a three-photon input state.} (a) Output probabilities for different choices of the delayed photon and of the measured output state contribution. Case I: Three-photon interference $P^{\mathrm{q}}_{i,j,k}$. Case II: distinguishable photon on input port 2 delayed (two-photon interference) $P^{(2)}_{i,j,k}$. Case III: identical photon on input port 3 delayed (two-photon interference) ${P^{(2)}}^{\prime}_{i,j,k}$. Blue bars: theoretical prediction obtained from the reconstructed tritter matrix $\mathcal{U}^r$ and partially distinguishable photon-pairs. Red bars: experimental results. (b) Visibilities $V=(\Gamma_{\infty}-\Gamma_{0})/\Gamma_{\infty}$ for different choices of the delayed photon and of the measured output state contribution. Here, $\Gamma_{\infty}$ and $\Gamma_{0}$ correspond respectively to the number of four-fold coincidences when the delayed photon is respectively out or inside the interference region. Case A: Distinguishable photon on input port 2 delayed (three-photon interference). Case B: identical photon on input port 3 delayed (three-photon interference). Case C: distinguishable photon on input port 2 out of interference region and identical photon on input port 3 delayed (two-photon interference). Blue bars: theoretical prediction obtained from the reconstructed tritter matrix $\mathcal{U}^r$ and partially distinguishable photon-pairs. Red bars: experimental results. Cyan bars: expected visibilities with classical states. Error bars are due to poissonian statistics of the measured signal.}
\label{fig:peak_measure}
\end{figure}

\textsl{\textbf{Experimental three photon interference}}. - We experimentally tested the three-photon coalescence by adopting the experimental setup shown in Fig. \ref{fig:chip_exp}. Four photons are produced by parametric down conversion: one of them acts as the trigger for coincidence detection, while the other three are coupled inside the tritter, after passing through different delay lines. The output modes are detected by using optical fiber beam splitters and single-photon avalanche photodiodes. Coincidences between different detectors allow us to reconstruct by post selection the probability of obtaining a given output state. By changing the relative delays between the three photons it is  possible to observe the coalescence effects in the different experimental conditions, as shown in Figs. \ref{fig:oumandel} (a-c). The tritter device, first characterized by single and two-photon states at wavelength $795$ nm (see Fig. \ref{fig:reconstruction}) has then been tested by three photons at wavelength $785$ nm, due to availability of higher transmissivity spectral filters at that wavelength. It has to be noted that the unitary matrix of the tritter $\mathcal{U}^{r}$ at $785$ nm is still in very good agreement with that of an ideal tritter (see Eq. \ref{eq:tritter_ideal}), yielding a similarity $\mathcal{S}_{r,t}^{\prime}=0.9595 \pm 0.0003$ (see Supplementary Note 3A and Supplementary Figure S4).  We have injected a three photon state $\ket{1,1,1}$ and observed the different output contributions $\ket{1,1,1},\ket{2,1,0},\ket{3,0,0}$. Due to the distinguishability between photons belonging to different pairs the expected visibilities of peaks (or dips depending on the observed contributions) change depending on which photon is delayed. We compare in Fig. \ref{fig:peak_measure} (a) the experimental and theoretical output states probabilities, obtained by delaying photons in different input modes for each measured output state. The expected probability values have been calculated by introducing in the theoretical model the distinguishability parameter $p$, which quantifies the distinguishability between the two different pairs, and by including six-photon events generated by the source. We investigated in Fig. \ref{fig:peak_measure} (b) three-photon interference when delaying the partially distinguishable photon [case A in the figure] or when delaying one of the two indistinguishable ones [case B in the figure]. In addition, we also considered the case when the partially distinguishable photon is kept out of the interference region and the two indistinguishable ones have a relative delay [case C in the figure]. The latter case corresponds to two-photon interference although three photons are input and measured. It is observed that cases A and B are clearly different from case C. This demonstrates that the output states that we observe in cases A and B are indeed due to three-photon interference and cannot be attributed to two-photon interaction. In order to characterize whether the measured three-photon interference can be attributed to a quantum effect, we also compared the obtained visibilities with the classical bounds, calculated by injecting the tritter with three equal-amplitude coherent states with randomized phases \cite{Metc12}. We observe that the results obtained with the $\vert 1,1,1 \rangle$ input state outperform classical predictions, thus demonstrating the quantum origin of the measured three-photon interference. Details on the data analysis can be found in the Supplementary Note 4B.

%
%

\section{Discussion}

We reported on the experimental observation of a three-photon bosonic coalescence effect occurring within an integrated, 3-dimensional, 3-port beam splitter realized by the ultrafast laser-writing technique. To the best of our knowledge, this result represents the first experimental observation of the bosonic coalescence of three undistinguishable photons in a symmetric three-port device. The intrinsically stable and compact nature of this device makes it a fundamental building block for future complex networks implemented by integrated optical circuits. These devices are expected to open new perspectives in many research areas of quantum information. Indeed, the $N$-port beam splitters may find a wide range of applications in both quantum interferometry and quantum metrology. A first theoretical analysis performed in Ref. \cite{Spag12}, shows that these devices can be exploited to build $N$-mode interferometers where the adoption of quantum input states can lead to a significant enhancement in phase estimation protocols. Furthermore, the capability of conditionally generating path-entangled states may be increased by adopting structures composed of $N$-port beam splitters rather than conventional two-mode couplers \cite{Pryd03}. This is expected to lead to a corresponding reduction in complexity of the experimental schemes in terms of the number of optical elements. 

Several other contexts may benefit from the adoption of $N$-port beam splitters. For instance, they can be adopted for the realization of ``proof-of-principle'' quantum simulators \cite{Lloy96}, and, combined in a modular structure, they can be used to realize full-scale quantum simulators for large size quantum systems. The adoption of multiphoton-multimode platforms may indeed disclose the ``hard-to-simulate'' scenario by adopting linear optics to implement a computational power beyond the one of a classical computer \cite{Aaro10}. Furthermore, the development of complex integrated quantum devices may lead to implement fundamental tests of quantum mechanics, such as nonlocality tests, for quantum systems of increasing dimensionality \cite{Gruc11}.

\section*{Methods}

\textsl{\textbf{Waveguides fabrication}}.- Waveguides are directly written in borosilicate glass substrate (EAGLE2000, Corning Inc.) using a cavity-dumped Yb:KYW oscillator, which provides 300 fs pulses at 1 MHz repetition rate. The laser beam is focused, by a 0.6-NA 50$\times$ microscope objective, at 170 $\mu$m (average) depth under the glass surface. Optimum irradiation parameters for single-mode waveguiding at 795 nm wavelength are 220 nJ pulse energy and 40 mm/s translation speed. Sample translation at uniform speed is provided by high precision, three-axis air-bearing stages (Aerotech FiberGlide3D, Aerotech Inc). The produced tritter is subsequently pigtailed by bonding fiber arrays on both sides of the device in order to obtain a compact and portable device. 

\textsl{\textbf{Four-photon source and detection apparatus}}.- Let us refer to the experimental setup shown in Fig. \ref{fig:chip_exp} in the main text. A $2$ mm thick $\beta$-barium borate crystal (BBO) cut for type-II phase matching, is pumped by the second harmonic of a Ti:Sa mode-locked laser beam, and generates via spontaneous parametric down conversion at second order two polarization photon pairs over two different spatial modes. The four-photons \cite{Eibl03,Radm09a} of the generated state are spectrally filtered by interferential filters (wavelength $\lambda = 785$ nm, $\Delta \lambda = 3$ nm) and selected by four single mode fibers placed at the output of two polarizing beam splitters. The spatial and temporal walk-off is compensated by inserting a waveplate and a $1$ mm thick BBO crystal on each spatial mode. One of the four photons is detected to work as the trigger of the experiment. The other three photons are injected each into a single mode fiber and, after passing through three delay lines and after polarization compensation, are coupled inside the tritter device. The output modes are sent to the detection stage. Four-fold coincidence detections are measured. The $\vert 1,1,1 \rangle$ contribution is measured by directly sending each output mode to a single photon detector. The $\vert 2,1,0 \rangle$ ($\vert 3,0,0 \rangle$) contribution is measured by splitting mode $1$ in two (three) equal parts by means of fiber beam-splitters and by placing a single photon detector on each of the two (three) outputs of the fiber beam-splitter system. The efficiencies for the three schemes including the trigger detector are respectively $\eta^{4}$, $(2/3)\eta^{4}$ and $(2/9) \eta^{4}$, being $\eta$ the quantum efficiency of a single detector.


\begin{thebibliography}{10}
\expandafter\ifx\csname url\endcsname\relax
  \def\url#1{\texttt{#1}}\fi
\expandafter\ifx\csname urlprefix\endcsname\relax\def\urlprefix{URL }\fi
\providecommand{\bibinfo}[2]{#2}
\providecommand{\eprint}[2][]{\url{#2}}

\bibitem{Hong87}
\bibinfo{author}{Hong, C.~K.}, \bibinfo{author}{Ou, Z.} \&
  \bibinfo{author}{Mandel, L.}
\newblock \bibinfo{title}{Measurement of subpicosecond time intervals between
  two photons by interference}.
\newblock \emph{\bibinfo{journal}{Phys. Rev. Lett.}}
  \textbf{\bibinfo{volume}{59}}, \bibinfo{pages}{2044-2046} (\bibinfo{year}{1987}).

\bibitem{Ou06}
\bibinfo{author}{Ou, Z.}
\newblock \bibinfo{title}{Temporal distinguishability of an n-photon state and
  its characterization by quantum interference}.
\newblock \emph{\bibinfo{journal}{Phys. Rev. A}} \textbf{\bibinfo{volume}{74}},
  \bibinfo{pages}{063808} (\bibinfo{year}{2006}).

\bibitem{Yama03}
\bibinfo{author}{Yamamoto, T.}, \bibinfo{author}{Koashi, M.},
  \bibinfo{author}{Ozdemir, S.} \& \bibinfo{author}{Imoto, N.}
\newblock \bibinfo{title}{Experimental extraction of an entangled photon pair
  from two identically decohered pairs}.
\newblock \emph{\bibinfo{journal}{Nature (London)}}
  \textbf{\bibinfo{volume}{421}}, \bibinfo{pages}{343-346} (\bibinfo{year}{2003}).

\bibitem{Walt04}
\bibinfo{author}{Walther, P.} \emph{et~al.}
\newblock \bibinfo{title}{De broglie wavelength of a non-local four-photon
  state}.
\newblock \emph{\bibinfo{journal}{Nature (London)}}
  \textbf{\bibinfo{volume}{429}}, \bibinfo{pages}{158-161} (\bibinfo{year}{2004}).

\bibitem{Kok07}
\bibinfo{author}{Kok, P.} \emph{et~al.}
\newblock \bibinfo{title}{Linear optical quantum computing with photonic
  qubits}.
\newblock \emph{\bibinfo{journal}{Rev. Mod. Phys.}}
  \textbf{\bibinfo{volume}{79}}, \bibinfo{pages}{135-174} (\bibinfo{year}{2007}).

\bibitem{Knil00}
\bibinfo{author}{Knill, E.}, \bibinfo{author}{Laflamme, R.} \&
  \bibinfo{author}{Milburn, G.~J.}
\newblock \bibinfo{title}{A scheme for efficient quantum computation with
  linear optics}.
\newblock \emph{\bibinfo{journal}{Nature (London)}}
  \textbf{\bibinfo{volume}{409}}, \bibinfo{pages}{46-52} (\bibinfo{year}{2000}).

\bibitem{Shih88}
\bibinfo{author}{Shih, Y.~H.} \& \bibinfo{author}{Alley, C.~O.}
\newblock \bibinfo{title}{New type of einstein-podolsky-rosen-bohm experiment
  using pairs of light quanta produced by optical parametric down conversion}.
\newblock \emph{\bibinfo{journal}{Phys. Rev. Lett.}}
  \textbf{\bibinfo{volume}{61}}, \bibinfo{pages}{2921-2924} (\bibinfo{year}{1988}).

\bibitem{XuBo02}
\bibinfo{author}{Zou, X.}, \bibinfo{author}{Pahlke, K.} \&
  \bibinfo{author}{Mathis, W.}
\newblock \bibinfo{title}{Generation of entangled photon states by using linear
  optical elements}.
\newblock \emph{\bibinfo{journal}{Phys. Rev. A}} \textbf{\bibinfo{volume}{66}},
  \bibinfo{pages}{014102} (\bibinfo{year}{2002}).

\bibitem{Gree93}
\bibinfo{author}{Greenberger, D.~M.}, \bibinfo{author}{Home, M.~A.} \&
  \bibinfo{author}{Zeilinger, A.}
\newblock \bibinfo{title}{Multiparticle interferometry and the quantum
  superposition principle}.
\newblock \emph{\bibinfo{journal}{Phys. Today}} \textbf{\bibinfo{volume}{46}},
  \bibinfo{pages}{22-29} (\bibinfo{year}{1993}).

\bibitem{Gree00}
\bibinfo{author}{Greenberger, D.~M.}, \bibinfo{author}{Home, M.~A.} \&
  \bibinfo{author}{Zeilinger, A.}
\newblock \bibinfo{title}{Similarities and differences between two-particle and
  three-particle interference}.
\newblock \emph{\bibinfo{journal}{Fortschr. Phys.}}
  \textbf{\bibinfo{volume}{48}}, \bibinfo{pages}{243-252} (\bibinfo{year}{2000}).

\bibitem{Pan12}
\bibinfo{author}{Pan, J.-W.} \emph{et~al.}
\newblock \bibinfo{title}{Multiphoton entanglement and interferometry}.
\newblock \emph{\bibinfo{journal}{Rev. Mod. Phys.}}
  \textbf{\bibinfo{volume}{84}}, \bibinfo{pages}{777-838} (\bibinfo{year}{2012}).

\bibitem{Zuko97}
\bibinfo{author}{Zukowski, M.}, \bibinfo{author}{Zeilinger, A.} \&
  \bibinfo{author}{Horne, M.~A.}
\newblock \bibinfo{title}{Realizable higher-dimensional two-particle
  entanglements via multiport beam splitters}.
\newblock \emph{\bibinfo{journal}{Phys. Rev. A}} \textbf{\bibinfo{volume}{55}},
  \bibinfo{pages}{2564-2579} (\bibinfo{year}{1997}).

\bibitem{Camp00}
\bibinfo{author}{Campos, R.~A.}
\newblock \bibinfo{title}{Three-photon hong-ou-mandel interference at a
  multiport mixer}.
\newblock \emph{\bibinfo{journal}{Phys. Rev. A}} \textbf{\bibinfo{volume}{62}},
  \bibinfo{pages}{013809} (\bibinfo{year}{2000}).

\bibitem{Howe00}
\bibinfo{author}{Howell, J.~C.} \& \bibinfo{author}{Yeazell, J.~A.}
\newblock \bibinfo{title}{Quantum computation through entangling single photons
  in multipath interferometers}.
\newblock \emph{\bibinfo{journal}{Phys. Rev. Lett.}}
  \textbf{\bibinfo{volume}{85}}, \bibinfo{pages}{198-201} (\bibinfo{year}{2000}).

\bibitem{Reck94}
\bibinfo{author}{Reck, M.}, \bibinfo{author}{Zeilinger, A.},
  \bibinfo{author}{Bernstein, H.~J.} \& \bibinfo{author}{Bertani, P.}
\newblock \bibinfo{title}{Experimental realization of any discrete unitary
  operator}.
\newblock \emph{\bibinfo{journal}{Phys. Rev. Lett.}}
  \textbf{\bibinfo{volume}{73}}, \bibinfo{pages}{58-61} (\bibinfo{year}{1994}).

\bibitem{Tich10}
\bibinfo{author}{Tichy, M.~C.}, \bibinfo{author}{Tiersch, M.},
  \bibinfo{author}{{de Melo}, F.}, \bibinfo{author}{Mintert, F.} \&
  \bibinfo{author}{Buchleitner, A.}
\newblock \bibinfo{title}{Zero-transmission law for multiport beam splitters}.
\newblock \emph{\bibinfo{journal}{Phys. Rev. Lett.}}
  \textbf{\bibinfo{volume}{104}}, \bibinfo{pages}{220405}
  (\bibinfo{year}{2010}).

\bibitem{Spag12}
\bibinfo{author}{Spagnolo, N.} \emph{et~al.}
\newblock \bibinfo{title}{Quantum interferometry with 3-dimensional geometry}.
\newblock \emph{\bibinfo{journal}{Sci. Rep.}} \textbf{\bibinfo{volume}{2}},
  \bibinfo{pages}{862} (\bibinfo{year}{2012}).

\bibitem{Weih96}
\bibinfo{author}{Weihs, G.}, \bibinfo{author}{Reck, M.},
  \bibinfo{author}{Weinfurter, H.} \& \bibinfo{author}{Zeilinger, A.}
\newblock \bibinfo{title}{Two-photon interference in optical fiber multiports}.
\newblock \emph{\bibinfo{journal}{Phys. Rev. A}} \textbf{\bibinfo{volume}{54}},
  \bibinfo{pages}{893-897} (\bibinfo{year}{1996}).

\bibitem{Kowa05}
\bibinfo{author}{Kowalevicz, A.~M.} \emph{et~al.}
\newblock \bibinfo{title}{Three-dimensional photonic devices fabricated in
  glass by use of a femtosecond laser oscillator}.
\newblock \emph{\bibinfo{journal}{Opt. Letters}}
  \textbf{\bibinfo{volume}{30}}, \bibinfo{pages}{1060-1062} (\bibinfo{year}{2005}).

\bibitem{Suzu06}
\bibinfo{author}{Suzuki, K.}, \bibinfo{author}{Sharma, V.},
  \bibinfo{author}{Fujimoto, J.~G.}, \bibinfo{author}{Ippen, E.~P.} \&
  \bibinfo{author}{Nasu, Y.}
\newblock \bibinfo{title}{Characterization of symmetric [3 x 3] directional
  couplers fabricated by direct writing with a femtosecond laser oscillator}.
\newblock \emph{\bibinfo{journal}{Phys. Rev. A}} \textbf{\bibinfo{volume}{80}},
  \bibinfo{pages}{032318} (\bibinfo{year}{2009}).

\bibitem{With12}
\bibinfo{author}{Meany, T.} \emph{et~al.}
\newblock \bibinfo{title}{Non-classical interference in integrated 3d
  multiports}.
\newblock \emph{\bibinfo{journal}{Opt. Express}} \textbf{\bibinfo{volume}{20}},
  \bibinfo{pages}{26895-26905} (\bibinfo{year}{2012}).

\bibitem{Gatt08}
\bibinfo{author}{Gattass, R.~R.} \& \bibinfo{author}{Mazur, E.}
\newblock \bibinfo{title}{Femtosecond laser micromachining in transparent
  materials}.
\newblock \emph{\bibinfo{journal}{Nat. Photon.}}
  \textbf{\bibinfo{volume}{2}}, \bibinfo{pages}{219-225} (\bibinfo{year}{2008}).

\bibitem{Dell09}
\bibinfo{author}{{Della Valle}, G.}, \bibinfo{author}{Osellame, R.} \&
  \bibinfo{author}{Laporta, P.}
\newblock \bibinfo{title}{Micromachining of photonic devices by femtosecond
  laser pulses}.
\newblock \emph{\bibinfo{journal}{J. Opt. A}}
  \textbf{\bibinfo{volume}{11}}, \bibinfo{pages}{013001}
  (\bibinfo{year}{2009}).

\bibitem{Osel11}
\bibinfo{author}{Osellame, R.}, \bibinfo{author}{Hoekstra, H. J. W.~M.},
  \bibinfo{author}{Cerullo, G.} \& \bibinfo{author}{Pollnau, M.}
\newblock \bibinfo{title}{Femtosecond laser microstructuring: an enabling tool
  for optofluidic lab-on-chips}.
\newblock \emph{\bibinfo{journal}{Laser Phot. Rev.}}
  \textbf{\bibinfo{volume}{5}}, \bibinfo{pages}{442-463} (\bibinfo{year}{2011}).

\bibitem{Cres11}
\bibinfo{author}{Crespi, A.} \emph{et~al.}
\newblock \bibinfo{title}{Integrated photonics quantum gates for polarization
  qubits}.
\newblock \emph{\bibinfo{journal}{Nat. Comm.}}
  \textbf{\bibinfo{volume}{2}}, \bibinfo{pages}{566} (\bibinfo{year}{2011}).

\bibitem{Owen11}
\bibinfo{author}{Owens, J.~O.} \emph{et~al.}
\newblock \bibinfo{title}{Two-photon quantum walks in an elliptical
  direct-write waveguide array}.
\newblock \emph{\bibinfo{journal}{New J. Phys.}}
  \textbf{\bibinfo{volume}{13}} \bibinfo{pages}{075003} (\bibinfo{year}{2011}).

\bibitem{Sans12}
\bibinfo{author}{Sansoni, L.} \emph{et~al.}
\newblock \bibinfo{title}{Two-particle bosonic-fermionic quantum walk via
  integrated photonics}.
\newblock \emph{\bibinfo{journal}{Phys. Rev. Lett.}}
  \textbf{\bibinfo{volume}{108}}, \bibinfo{pages}{010502}
  (\bibinfo{year}{2012}).

\bibitem{Peru11}
\bibinfo{author}{Peruzzo, A.}, \bibinfo{author}{Laing, A.},
  \bibinfo{author}{Politi, A.}, \bibinfo{author}{Rudolph, T.} \&
  \bibinfo{author}{O'Brien, J.~L.}
\newblock \bibinfo{title}{Multimode quantum interference of photons in
  multiport integrated devices}.
\newblock \emph{\bibinfo{journal}{Nat. Comm.}}
  \textbf{\bibinfo{volume}{2}}, \bibinfo{pages}{244} (\bibinfo{year}{2011}).

\bibitem{Metc12}
\bibinfo{author}{Metcalf, B.~J.} \emph{et~al.}
\newblock \bibinfo{title}{Multi-photon quantum interference in a multi-port
  integrated photonic device}.
\newblock \emph{\bibinfo{journal}{Nat. Comm.}}
  \textbf{\bibinfo{volume}{4}}, \bibinfo{pages}{1356} (\bibinfo{year}{2013}).

\bibitem{Pryd03}
\bibinfo{author}{Pryde, G.~J.} \& \bibinfo{author}{White, A.~G.}
\newblock \bibinfo{title}{Creation of maximally entangled photon-number states
  using optical fiber multiports}.
\newblock \emph{\bibinfo{journal}{Phys. Rev. A}} \textbf{\bibinfo{volume}{68}},
  \bibinfo{pages}{052315} (\bibinfo{year}{2003}).

\bibitem{Lloy96}
\bibinfo{author}{Lloyd, S.}
\newblock \bibinfo{title}{Universal quantum simulators}.
\newblock \emph{\bibinfo{journal}{Science}} \textbf{\bibinfo{volume}{273}},
  \bibinfo{pages}{1073-1078} (\bibinfo{year}{1996}).

\bibitem{Aaro10}
\bibinfo{author}{Aaronson, S.} \& \bibinfo{author}{Arkhipov, A.}
\newblock \bibinfo{title}{The computational complexity of linear optics}.
\newblock Preprint at http://arxiv.org/abs/1011.3245 (2010).

\bibitem{Gruc11}
\bibinfo{author}{Gruca, J.}, \bibinfo{author}{Laskowski, W.} \&
  \bibinfo{author}{Zukowski, M.}
\newblock \bibinfo{title}{Nonclassicality of pure two-qutrit entangled states}.
\newblock \emph{\bibinfo{journal}{Phys. Rev. A}} \textbf{\bibinfo{volume}{85}},
  \bibinfo{pages}{022118} (\bibinfo{year}{2012}).

\bibitem{Eibl03}
\bibinfo{author}{Eibl, M.} \emph{et~al.}
\newblock \bibinfo{title}{Experimental observation of four-photon entanglement
  from parametric down-conversion}.
\newblock \emph{\bibinfo{journal}{Phys. Rev. Lett.}}
  \textbf{\bibinfo{volume}{90}}, \bibinfo{pages}{200403}
  (\bibinfo{year}{2003}).

\bibitem{Radm09a}
\bibinfo{author}{Radmark, M.}, \bibinfo{author}{Zukowski, M.} \&
  \bibinfo{author}{Bourennane, M.}
\newblock \bibinfo{title}{Experimental test of fidelity limits in six-photon
  interferometry and of rotational invariance properties of the photonic
  six-qubit entanglement singlet state}.
\newblock \emph{\bibinfo{journal}{Phys. Rev. Lett.}}
  \textbf{\bibinfo{volume}{103}}, \bibinfo{pages}{150501}
  (\bibinfo{year}{2009}).

\end{thebibliography}

\section{Acknowledgements}
This work was supported by FIRB-Futuro in Ricerca HYTEQ and ERC (European Research Council)—Starting Grant 3D-QUEST (3D-Quantum Integrated Optical Simulation; grant agreement no. 307783): http://www.3dquest.eu.

\section{Author contributions}
All the authors coinceived the experiment. A.C., R.R., R.O. designed and fabricated the integrated tritter chip and performed the characterization measurements in the classical regime. N.S., C.V., L.A., F.S., P.M. developed the quantum theory underlying the experiment and carried out the measurements in the quantum regime. All authors discussed the results and participated in the manuscript preparation.

\section{Additional Informations}
\textbf{Competing financial interests:} The authors declare no competing financial interests.

\end{document}


\title{Three-photon bosonic coalescence in an integrated tritter -\\Supplementary Information}

\author{Nicol\`o Spagnolo}
\affiliation{Dipartimento di Fisica, Sapienza Universit\`{a} di Roma, Piazzale Aldo Moro 5, I-00185 Roma, Italy}
\author{Chiara Vitelli}
\affiliation{Center of Life NanoScience @ La Sapienza, Istituto Italiano di Tecnologia, Viale Regina Elena, 255, I-00185 Roma, Italy}
\affiliation{Dipartimento di Fisica, Sapienza Universit\`{a} di Roma, Piazzale Aldo Moro 5, I-00185 Roma, Italy}
\author{Lorenzo Aparo}
\affiliation{Dipartimento di Fisica, Sapienza Universit\`{a} di Roma, Piazzale Aldo Moro 5, I-00185 Roma, Italy}
\author{Paolo Mataloni}
\affiliation{Dipartimento di Fisica, Sapienza Universit\`{a} di Roma, Piazzale Aldo Moro 5, I-00185 Roma, Italy}
\author{Fabio Sciarrino}
\email{fabio.sciarrino@uniroma1.it}
\affiliation{Dipartimento di Fisica, Sapienza Universit\`{a} di Roma, Piazzale Aldo Moro 5, I-00185 Roma, Italy}
\author{Andrea Crespi}
\affiliation{Istituto di Fotonica e Nanotecnologie, Consiglio Nazionale delle Ricerche (IFN-CNR), Piazza Leonardo da Vinci, 32, I-20133 Milano, Italy}
\affiliation{Dipartimento di Fisica, Politecnico di Milano, Piazza Leonardo da Vinci, 32, I-20133 Milano, Italy}
\author{Roberta Ramponi}
\affiliation{Istituto di Fotonica e Nanotecnologie, Consiglio Nazionale delle Ricerche (IFN-CNR), Piazza Leonardo da Vinci, 32, I-20133 Milano, Italy}
\affiliation{Dipartimento di Fisica, Politecnico di Milano, Piazza Leonardo da Vinci, 32, I-20133 Milano, Italy}
\author{Roberto Osellame}
\email{roberto.osellame@polimi.it}
\affiliation{Istituto di Fotonica e Nanotecnologie, Consiglio Nazionale delle Ricerche (IFN-CNR), Piazza Leonardo da Vinci, 32, I-20133 Milano, Italy}
\affiliation{Dipartimento di Fisica, Politecnico di Milano, Piazza Leonardo da Vinci, 32, I-20133 Milano, Italy}

\maketitle

\newpage	

\begin{figure}[ht!]
\centering
\includegraphics[width=10cm]{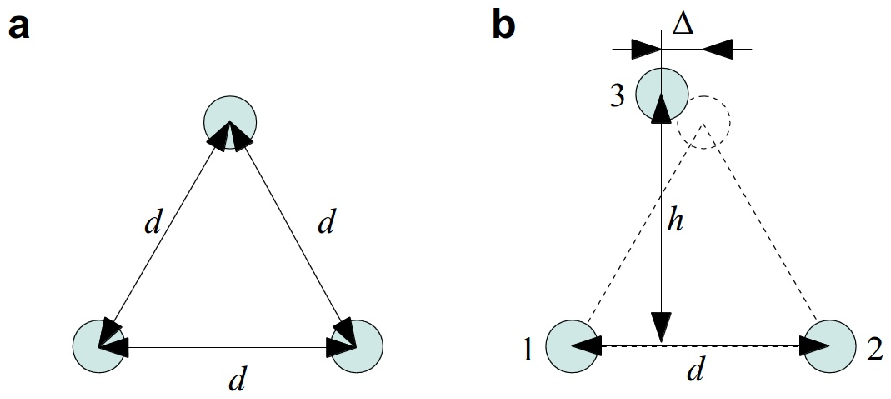}
\caption{}
\label{fig:crossSect}
\end{figure}
\vspace{-1.cm}
\noindent {\bf Supplementary Figure S1: Cross-section of an integrated tritter.} (a) Cross-section of an ideal three-waveguide coupler. The waveguides are disposed on the vertices of an equilateral triangle, with reciprocal distance $d$. (b) Cross-section of the three-waveguide coupler (tritter) realized in this work. To compensate for various non-idealities of the waveguide modes and fabrication process, the position of the third waveguide is tuned in the two directions. The waveguides constitute the vertices of a scalene triangle; the ideal equilateral geometry is slightly deformed.

\vspace{1.cm}

\begin{figure}[ht!]
\centering
\subfigure[]{\includegraphics[width=0.97\textwidth]{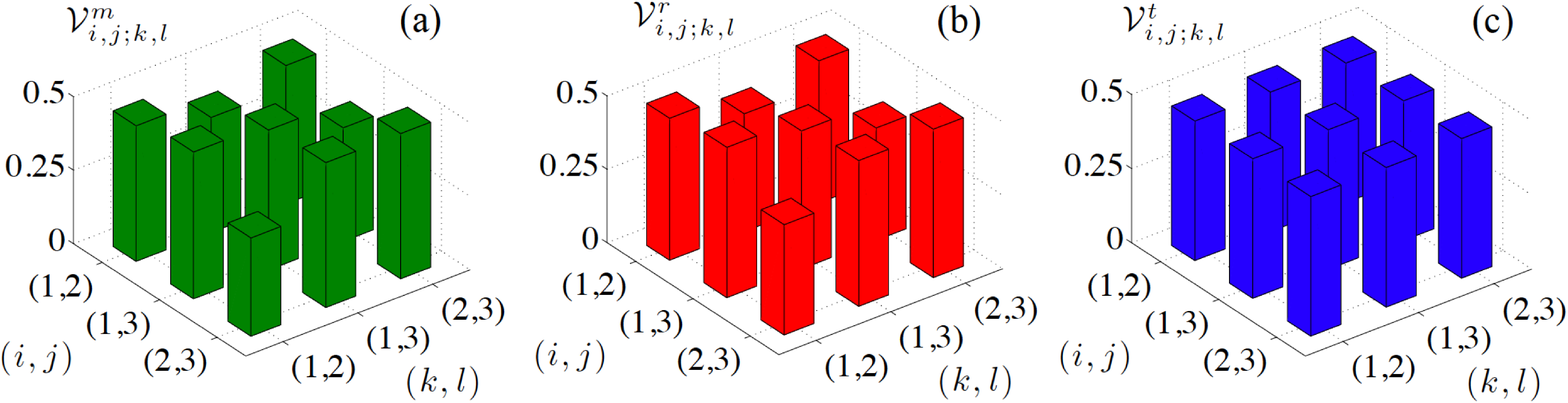}}
\caption{}
\label{fig:visibilities_U}
\end{figure}
\vspace{-1.cm}
\noindent {\bf Supplementary Figure S2: Two-photon Hong-Ou-Mandel interference at $795$ nm.} Visibilities of the dips at $795$ nm, for the input modes $(i,j)$ and the outputs $(k,l)$: (a) measured visibilities, (b) visibilities obtained with $\mathcal U^{r}$ and (c) visibilities expected with an ideal tritter $\mathcal{U}^{t}$.

\newpage

\begin{figure}[ht!]
\centering
\subfigure[]{\includegraphics[width=0.97\textwidth]{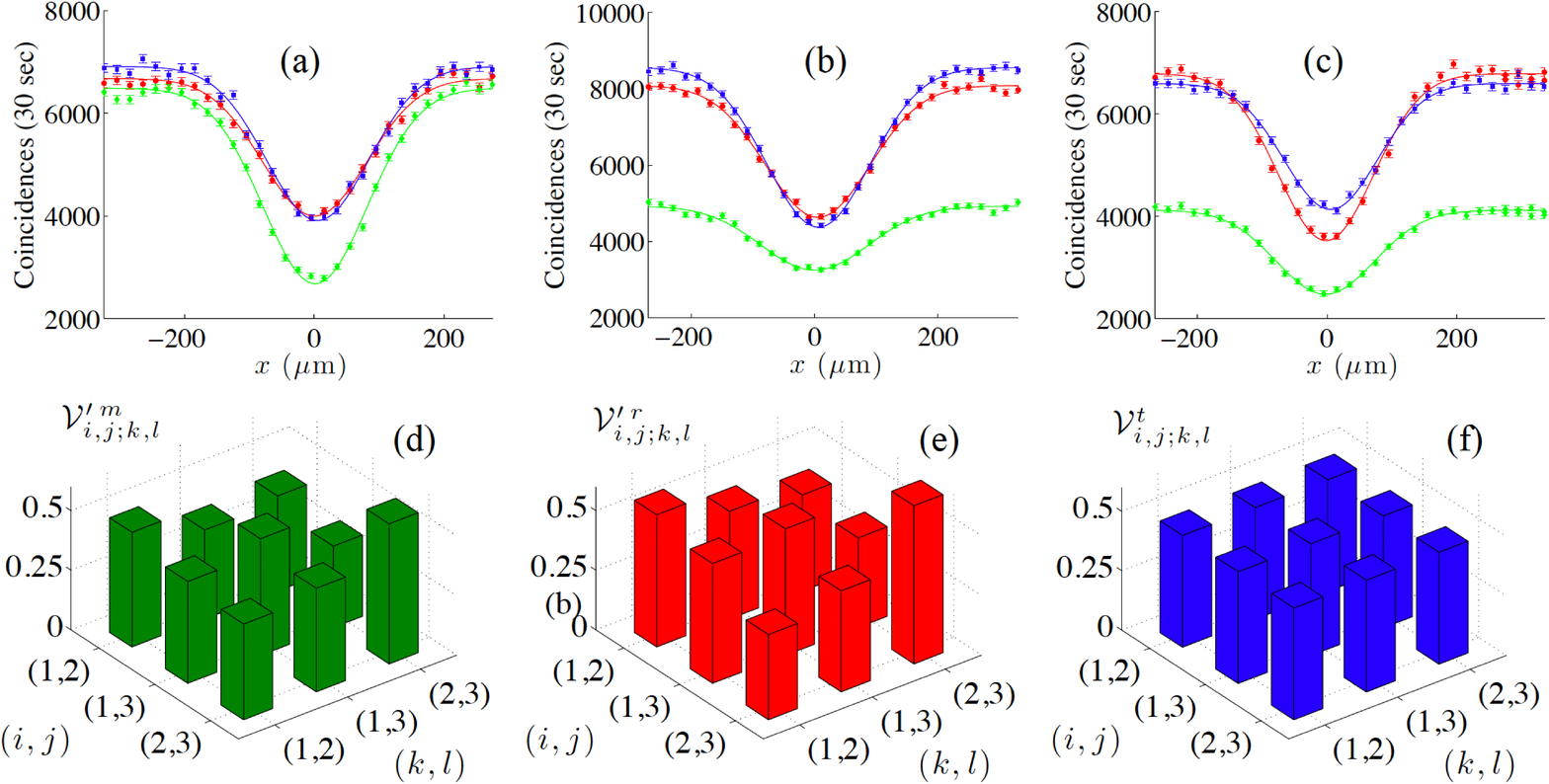}}
\caption{}
\label{fig:visibilities_U_785}
\end{figure}
\vspace{-1.cm}
\noindent {\bf Supplementary Figure S3: Two-photon Hong-Ou-Mandel interference at $785$ nm.} (a-c) Two-photon Hong-Ou-Mandel dips for input states: (a) $\vert 1,1,0 \rangle$, (b) $\vert 1,0,1 \rangle$ and (c) $\vert 0,1,1 \rangle$. Red points (experimental data) and red lines (best fit) correspond to the (1,1,0) output contribution. Blue points (experimental data) and blue lines (best fit) correspond to the (1,0,1) output contribution. Green points (experimental data) and green lines (best fit) correspond to the output (0,0,1) contribution. Error bars in the experimental points are due to poissonian statistics of the measured signal. (d-f) Corresponding visibilities of the dips, for the input modes $(i,j)$ and the outputs $(k,l)$: (d) measured visibilities, (e) visibilities obtained with $\mathcal{U}^{\prime \, r}$ and (f) visibilities expected with an ideal tritter $\mathcal{U}^{t}$.

\newpage

\begin{figure}[ht!]
\centering
\includegraphics[width=0.75\textwidth]{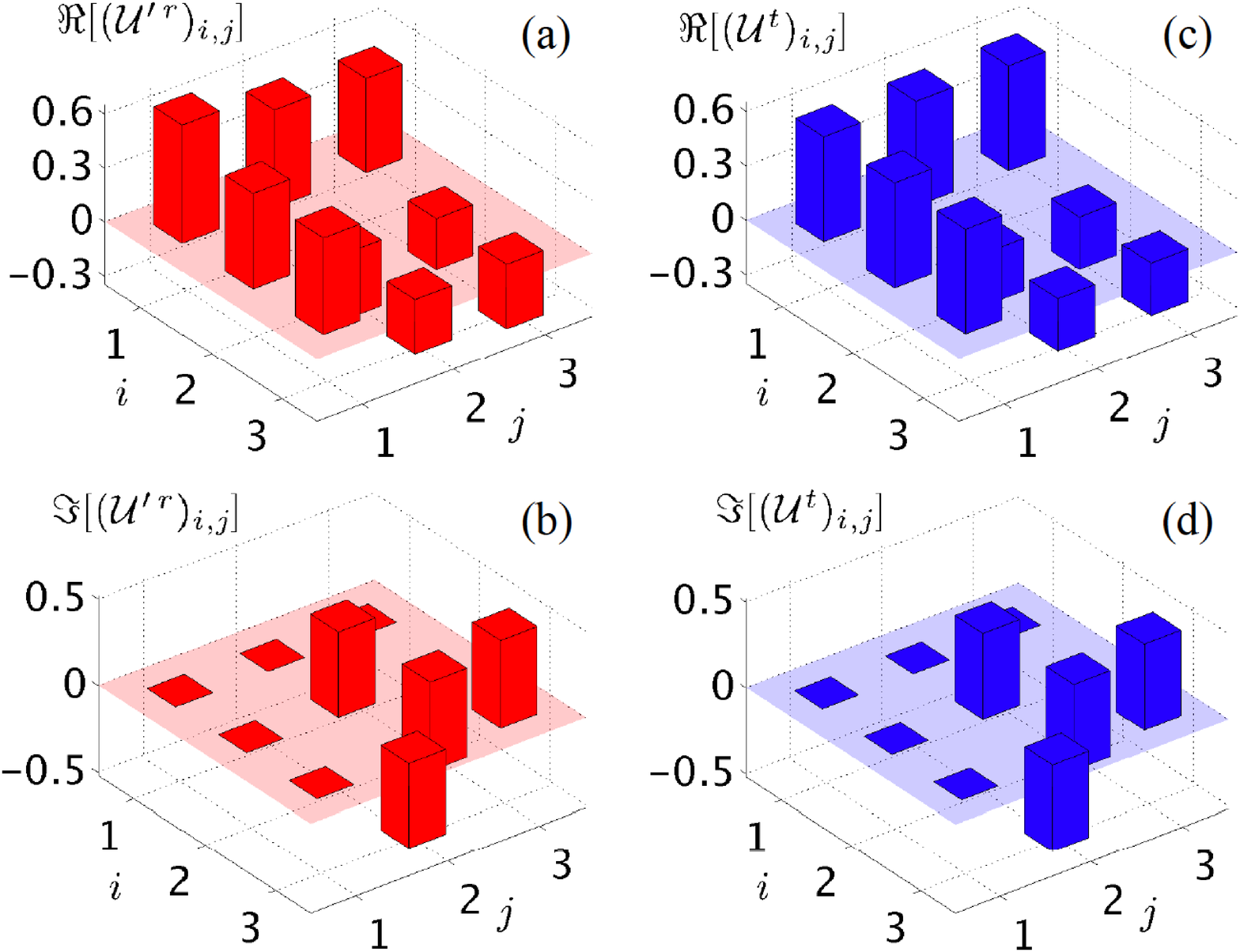}
\caption{}
\label{fig:reconstruction_785}
\end{figure}
\vspace{-1.cm}
\noindent {\bf Supplementary Figure S4: Reconstruction of the tritter matrix at $785$ nm.} Experimental reconstruction of the unitary matrix $\mathcal{U}^{\prime \, r}$ of the implemented tritter at $\lambda = 785$ nm compared with the theoretical one $\mathcal{U}^{t}$. (a)-(b) Real and imaginary parts of $\mathcal{U}^{\prime \, r}$. (c)-(d) Real and imaginary parts of $\mathcal{U}^{t}$.

\newpage

\begin{figure}[ht!]
\centering
\includegraphics[width=0.92\textwidth]{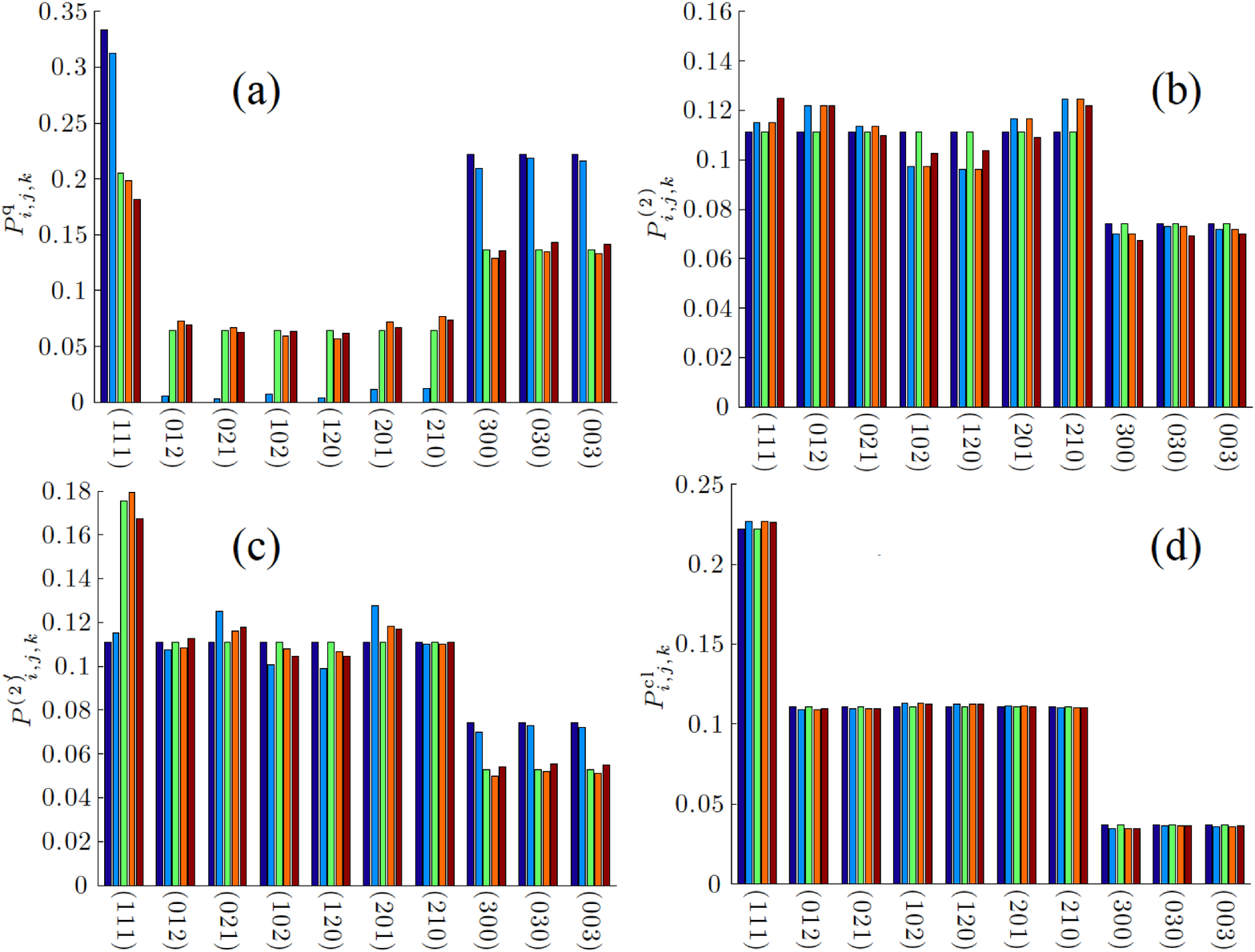}
\caption{}
\label{fig:prob}
\end{figure}
\vspace{-1.cm}
\noindent {\bf Supplementary Figure S5: Expected outcome probabilities for a three-photon input in state $\vert 1,1,1 \rangle$.} (a) Three simultaneous photons $P^{\mathrm{q}}_{i,j,k}$. (b) Distinguishable photon on input port $2$ out of interference region $P^{(2)}_{i,j,k}$. (c) Identical photon in input port $3$ out of interference region ${P^{(2)}}^{\prime}_{i,j,k}$. (d) All photons out of the interference region $P^{\mathrm{cl}}_{i,j,k}$. Blue bars: theoretical tritter matrix $\mathcal{U}^{t}$. Cyan bars: reconstructed tritter matrix $\mathcal{U}^{r}$. Green bars: theoretical tritter matrix $\mathcal{U}^{t}$ with photon indistinguishability $p=0.65$. Orange bars: reconstructed tritter matrix $\mathcal{U}^{r}$ with photon indistinguishability $p=0.65$. Red bars: reconstructed tritter matrix $\mathcal{U}^{r}$ with photon indistinguishability $p=0.65$ and six-photon contribution.

\newpage

\begin{table}[ht!]
\begin{center}
\begin{tabular}{|l|l|l|}
\hline
\multicolumn{3}{|c|}{Comparison quantum vs classical probability} \\
\hline
Input state & Output state & $P_{i,j,k}^{\mathrm{q}}/P_{i,j,k}^{\mathrm{cl}}$ \\ \hline
\multirow{3}{*}{$\ket{1,1,1}$} & $\ket{1,1,1}$ & $\;\;\;\;\;\;\;\;3/2$ \\
 & $\ket{3,0,0}$ & $\;\;\;\;\;\;\;\;\;6$ \\
 & $\ket{2,1,0}$ & $\;\;\;\;\;\;\;\;\;0$ \\
 \hline
\multirow{3}{*}{$\ket{2,1,0}$} & $\ket{1,1,1}$ & $\;\;\;\;\;\;\;\;\;0$ \\
 & $\ket{3,0,0}$ & $\;\;\;\;\;\;\;\;\;3$ \\
 & $\ket{2,1,0}$ & $\;\;\;\;\;\;\;\;\;1$ \\ \hline
\multirow{3}{*}{$\ket{3,0,0}$} & $\ket{1,1,1}$ & $\;\;\;\;\;\;\;\;\;1$ \\
 & $\ket{3,0,0}$ & $\;\;\;\;\;\;\;\;\;1$ \\
 & $\ket{2,1,0}$ & $\;\;\;\;\;\;\;\;\;1$ \\
\hline
\end{tabular}
\caption{}
\label{tab:S1}
\end{center}
\end{table}
\vspace{-1.cm}
{\bf Supplementary Table S1: Three-photon quantum and classical probabilities.} Calculation of the output probabilities for three-photon input states. $P^{\mathrm{q}}_{i,j,k}$ corresponds to three identical photons, while $P^{\mathrm{cl}}_{i,j,k}$ corresponds to three distinguishable photons.

\vspace{1.cm}

\begin{table}[ht!]
\begin{center}
\begin{tabular}{|l|l|l|}
\hline
\multicolumn{3}{|c|}{Comparison quantum vs classical probability} \\
\hline
Input state & Output state & $P_{i,j,k}^{\mathrm{q}}/P_{i,j,k}^{\mathrm{cl}}$ \\ \hline
\multirow{2}{*}{$\ket{1,1,0}$} & $\ket{1,1,0}$ & $\;\;\;\;\;\;\;\;1/2$ \\
 & $\ket{2,0,0}$ & $\;\;\;\;\;\;\;\;\;2$ \\
 \hline
\multirow{2}{*}{$\ket{2,0,0}$} & $\ket{1,1,0}$ & $\;\;\;\;\;\;\;\;\;1$ \\
 & $\ket{2,0,0}$ & $\;\;\;\;\;\;\;\;\;1$ \\
\hline
\end{tabular}
\caption{}
\label{tab:S2}
\end{center}
\end{table}
\vspace{-1.cm}
{\bf Supplementary Table S2: Two-photon quantum and classical probabilities.} Calculation of the output probabilities for two-photon input states. $P^{\mathrm{q}}_{i,j,k}$ corresponds to two identical photons, while $P^{\mathrm{cl}}_{i,j,k}$ corresponds to two distinguishable photons.

\newpage

\section*{Supplementary Note 1: Theoretical details}

\subsection*{A. Mathematical description of tritter}

\label{sec:tritterTeo}
The three-photon interference phenomena in a balanced integrated tritter can be modelled by considering the propagation of three photons in three equally-coupled waveguides. If $A_j^{\dag}$ are the creation operators for a photon in the $j$-th waveguide ($j = 1,2,3$), their evolution along the propagation coordinate $z$, in the Heisenberg picture, is described by the following equations [36]:
\begin{equation}
\left\lbrace
\begin{array}{l}
i \, \dfrac{d A_1^{\dag}}{d z} =  \beta A_1^{\dag} + k A_2^{\dag} + k A_3^{\dag}\\
i \, \dfrac{d A_2^{\dag}}{d z} =  k A_1^{\dag} + \beta A_2^{\dag} + k A_3^{\dag}\\
i \, \dfrac{d A_3^{\dag}}{d z} =  k A_1^{\dag} + k A_2^{\dag} + \beta A_3^{\dag}
\end{array}
\right.
\label{eq:tritterCoupledModes}
\end{equation}
where $\beta$ is the propagation constant and $k$ is the coupling coefficient between neighbouring waveguides (assumed constant for each couple of waveguides). 

Supplementary Eqs. \eqref{eq:tritterCoupledModes} admit an explicit solution as a function of $z$:
\begin{equation}
\begin{pmatrix} A^{\dag}_1  (z)\\ A^{\dag}_2  (z)\\A^{\dag}_3  (z)
\end{pmatrix}
=\mathcal{U} 
\begin{pmatrix}  A^{\dag}_1  (0)\\ A^{\dag}_2  (0)\\A^{\dag}_3  (0)
\end{pmatrix}
\label{eq:tritterGeneralSolution}
\end{equation}
with:
\begin{equation}
\mathcal{U} = \frac{1}{3} e^{- i \beta z} \begin{pmatrix}

2 e^{i \, k z} + e^{-i \, 2 k z} & e^{i \, 2 k z} - e^{i \, k z} & e^{i \, 2 k z} - e^{i \, k z} \\

 e^{i \, 2 k z} - e^{i \, k z} & 2 e^{i \, k z} + e^{-i \, 2 k z} & e^{i \, 2 k z} - e^{i \, k z} \\

e^{i \, 2 k z} - e^{i \, k z} & e^{i \, 2 k z} - e^{i \, k z} & 2 e^{i \, k z} + e^{-i \, 2 k z}

\end{pmatrix}
\label{eq:tritterGeneralMatrix}
\end{equation}
One can easily calculate that a balanced tritter, which distributes each input photon with equal probability to the three outputs, is obtained for a propagation length $z = L$ such that $k L = \frac{2}{9} \pi$. For compactness of notation we can define $a_j^{\dag} = A_j^{\dag} (0)$ as the creation operator of a photon at the $j$-th input port of the tritter and $b_j^{\dag} = A_j^{\dag} (L)$ as the creation operator of a photon at the $j$-th output port of the tritter. It holds:
\begin{equation}
\begin{pmatrix} b^{\dag}_1 \\ b^{\dag}_2 \\ b^{\dag}_3
\end{pmatrix}=\mathcal{U} \begin{pmatrix} a^{\dag}_1 \\ a^{\dag}_2 \\ a^{\dag}_3
\end{pmatrix}
\mbox{\:\:\:\:\:\: and \:\:\:\:\:\:}
\begin{pmatrix} a^{\dag}_1 \\ a^{\dag}_2 \\ a^{\dag}_3
\end{pmatrix}
=\mathcal{U}^{-1} \begin{pmatrix} b^{\dag}_1 \\ b^{\dag}_2 \\ b^{\dag}_3
\end{pmatrix}.
\end{equation}
with:
\begin{equation}
\mathcal{U} = \frac{1}{\sqrt{3}}  e^{- i \frac{2}{9} \frac{ \beta \pi }{k} + i \frac{4}{9} \pi} \begin{pmatrix}
1 & e^{\imath 2\pi/3} & e^{\imath 2\pi/3}\\ e^{\imath 2\pi/3} & 1 & e^{\imath 2\pi/3}\\ e^{\imath 2\pi/3} & e^{\imath 2\pi/3} & 1
\end{pmatrix}
\label{eq:BalancedTritterMatrix}
\end{equation}
This matrix turns out to be equivalent, up to multiplications for a phase factor of rows and columns (which correspond to putting a phase delay on the input and output ports of the tritter), to the real-bordered unitary matrix $\mathcal{U}^{t}$:
\begin{equation}
\label{eq:matrixteo}
\mathcal{U}^{t} = \frac{1}{\sqrt{3}} \begin{pmatrix} 1 & 1 & 1 \\ 1 & e^{\imath 2 \pi/3} & e^{\imath 4 \pi/3} \\ 1 & e^{\imath 4 \pi/3} & e^{\imath 8 \pi/3} \end{pmatrix}.
\end{equation}

\subsection*{B. Generalized bosonic coalescence}

By following the formalism introduced in Ref. [16] we now analyze the generalized coalescence effect in a configuration of $m$ ($m=2,3$) bosons prepared in the $3$ input ports of a tritter. 

Let us first consider different three-photon input states, and compare the expected output probabilities for outcomes $\vert i,j,k \rangle$ relative to a quantum $P^{\mathrm{q}}_{i,j,k}$ and a classical $P^{\mathrm{cl}}_{i,j,k}$ description. In particular, $P^{\mathrm{q}}_{i,j,k}$ corresponds to perfect interference between the three incident photons, while $P^{\mathrm{cl}}_{i,j,k}$ corresponds to the case where the three photons don't interfere and evolve indipendently in the device (see Supplementary Table S1). We observe that for the $\vert 1,1,1 \rangle$ and $\vert 2,1,0 \rangle$ input states three-photon quantum interferences is responsible for both enhancement and suppression of the output probabilities.

A similar result is obtained when two photons are impinging on the device. The comparison between the quantum and classical probabilities to obtain a given state at the exit of the three-port beam splitter is reported in Supplementary Table S2.

\subsection*{C. Multimode theory for three-photon bosonic coalescence} 

We now consider the case of three independent indistinguishable photons, each of them described by the wavepacket operator:
\begin{equation}
a_{i}^{\dag}(\tau_{i}) = \int_{-\infty}^{\infty} d \omega_{i} f_{i}(\omega_{i}) a_{i}^{\dag}(\omega_{i}) e^{- \imath \omega_{i} \tau_{i}},
\end{equation}
where $i=1,2,3$ labels the optical modes $\mathbf{k}_{i}$, $\tau_{i}$ is the time delay due to spatial propagation, $f_{i}(\omega_{i})$ is the spectral function of the photon wavepacket with central frequency $\omega_{0}$ and width $\delta \omega$:
\begin{equation}
f_{i}(\omega_{i}) = \frac{1}{(2 \pi \delta \omega^{2})^{1/4}} e^{-\frac{(\omega_{i}-\omega_{0})^{2}}{4 \delta \omega^{2}}}.
\end{equation}
The output state after propagation through a tritter is obtained by replacing the time evolution equations $a^{\dag}_{i}(\omega_{i}) = \sum_{i,j} (\mathcal{U}^{-1})_{ij} b^{\dag}_{j}(\omega_{i})$ in the input state:
\begin{equation}
\prod_{i=1}^{3} a^{\dag}_{i}(\tau_{i}) \vert 0 \rangle = \int_{-\infty}^{\infty} \int_{-\infty}^{\infty} \int_{-\infty}^{\infty} \prod_{i=1}^{3} \left( d \omega_{i} f_{i}(\omega_{i}) a^{\dag}_{i}(\omega_{i}) e^{-\imath \omega_{i} \tau_{i}}\right) \vert 0 \rangle,
\end{equation}
The probability $P_{1,1,1}$ of obtaining one photon on each output port is calculated by selecting the terms $b_{1}^{\dag}(\omega_{m})b_{2}^{\dag}(\omega_{n})b_{3}^{\dag}(\omega_{p})$ of the output state. For the symmetric tritter $\vert T \vert = \vert R \vert = 1/\sqrt{3}$, we obtain:
\begin{equation}
P_{1,1,1}(\tau_{1},\tau_{2},\tau_{3}) = \frac{1}{9} [2- e^{-\delta \omega^{2}(\tau_{1}-\tau_{2})^{2}} - e^{-\delta \omega^{2}(\tau_{1}-\tau_{3})^{2}} 
- e^{-\delta \omega^{2}(\tau_{2}-\tau_{3})^{2}} + 4 e^{-\delta \omega^{2}(\tau_{1}^{2}+\tau_{2}^{2}+\tau_{3}^{2}-\tau_{1} \tau_{2} - \tau_{1} \tau_{3} - \tau_{2} \tau_{3})} ].
\end{equation}
By an analogous procedure, the probabilities $P_{2,1,0}$ and $P_{3,0,0}$ of obtaining the outcomes $(2,1,0)$ and $(3,0,0)$ read:
\begin{equation}
P_{2,1,0}(\tau_{1},\tau_{2},\tau_{3})= \frac{1}{9}[1- e^{-\delta \omega^{2}(\tau_{1}^{2}+\tau_{2}^{2}+\tau_{3}^{2}-\tau_{1} \tau_{2} - \tau_{1} \tau_{3} - \tau_{2} \tau_{3})}],\\
\end{equation}
and:
\begin{equation}
P_{3,0,0}(\tau_{1},\tau_{2},\tau_{3})= \frac{1}{27} [1+ e^{-\delta \omega^{2}(\tau_{1}-\tau_{2})^{2}} + e^{-\delta \omega^{2}(\tau_{1}-\tau_{3})^{2}} + e^{-\delta \omega^{2}(\tau_{2}-\tau_{3})^{2}} + 2 e^{-\delta \omega^{2}(\tau_{1}^{2}+\tau_{2}^{2}+\tau_{3}^{2}-\tau_{1} \tau_{2} - \tau_{1} \tau_{3} - \tau_{2} \tau_{3})} ].
\end{equation}
We observe the presence of the two-photon interference terms $e^{-\delta \omega^{2}(\tau_{m}-\tau_{n})^{2}}$, depending only on the relative delay between input photons $m$ and $n$, and a pure three-photon interfence term $e^{-\delta \omega^{2}(\tau_{1}^{2}+\tau_{2}^{2}+\tau_{3}^{2}-\tau_{1} \tau_{2} - \tau_{1} \tau_{3} - \tau_{2} \tau_{3})}$, depending on the relative delay between the three input photons.

We conclude by discussing the effect of spectral mismatch between the photons in the Hong-Ou-Mandel output probabilities $P_{i,j,k}(\tau_{1},\tau_{2},\tau_{3})$. In this case, we can expect that the two-photon interference terms will be reduced by a factor describing the spectral overlap between the modes $m$ and $n$. Analogously, the three-photon interference term will be reduced by a factor describing the spectral overlap between the three modes $m$, $n$ and $p$.

\section*{Supplementary Note 2: Geometry optimization of the femtosecond laser written tritter}

In this section we will discuss how to design a three-waveguide coupler in order to achieve symmetric power repartition in the three arms, i.e. in order to realize a symmetric tritter.
If waveguides with perfectly circular guided modes were adopted, a symmetric coupling between the three waveguides would be achieved by placing them at the same relative distance $d$. In other words, within the coupling region (that is, the region in which they are close to each other and couple by evanescent field) the three waveguides should lay at the vertices of an equilateral triangle, in a three-dimensional layout [Supplementary Fig. S\ref{fig:crossSect} (a)]. 

As a matter of fact, the femtosecond laser written waveguides adopted in this work support elliptical guided modes [27]. This causes an angle-dependent coupling coefficient between waveguides [37]. In addition, other fabrication non-idealities may occur (see e.g. Ref. [20]) which may bring about further asymmetries in the coupling. One can infer that in this case a symmetric equilateral-triangle geometry would not provide symmetric coupling; indeed, in order to realize an accurately balanced tritter, it is crucial to develop a robust experimental procedure for optimizing the fabrication geometric parameters, which can compensate all the non-idealities.

As a first step we chose the separation $d$ between the two waveguides 1 and 2, which are at the same depth in the substrate [see Supplementary Fig. S\ref{fig:crossSect} (b)]. The coupling coefficient $k$ exponentially increases [37] with decreasing $d$: a small value of $d$ is preferable to obtain a compact device. In our case it was chosen $d=$11~$\mu$m, which allows to fabricate three waveguides with the three-dimensional layout of Supplementary Fig. S\ref{fig:crossSect} (b) without touching each other for every vertical distance $h$ of the third waveguide.

As discussed above, the choice of $d$ determines the value of $k$. As shown in Supplementary Note 1A, $L$ is then fixed by the choice of $d$: in particular, for a balanced tritter, it must be $k L = \frac{2}{9} \pi$. Actually, the interaction length $L$ can be experimentally optimized more conveniently by studying two-waveguide couplers, which are much simpler devices. For a two-waveguide coupler, equations similar to Supplementary Eqs. \eqref{eq:tritterCoupledModes} can be written [38] and the coupler reflectivity results to be: 
\begin{equation}
R = \sin^2 \left( k L \right).
\end{equation}
Here $R$ is defined as $R = \protect \frac {P^{\protect \mathrm{out}}_1}{P^{\protect \mathrm {out}}_1 + P^{\protect \mathrm {out}}_2}$, where $P^{\protect \mathrm {out}}_1$ is the output power from waveguide 1, $P^{\protect \mathrm {out}}_2$ is the output power from waveguide 2 and coherent light is injected in waveguide 1. We fabricated several devices consisting of the waveguides 1 and 2 at distance $d$ with different lengths $L$, but without the third waveguide, i.e. simple planar directional couplers. This allowed to retrieve the optimum interaction length, which must satisfy $R = \sin^2 \left( \frac{2}{9} \pi \right) \simeq 41 \%$. In our case it yielded $L \simeq 2.5$~mm. It should be reminded that this length, providing $41\%$ reflectivity in a two-waveguide coupler, corresponds to a balanced $33\%$ reflectivity in a three-waveguide tritter, as discussed above and demonstrated in Supplementary Note 1A.

As a following step, the position of the third waveguide must be optimized. For this purpose, it is important to note that the ellipticity of the guided modes, which gives an angle-dependent coupling, requires the distance between waveguides 1 and 2 to be different with respect to the distances between those waveguides and waveguide 3. The former \textit{equilateral} triangle, at whose vertices the waveguides should lay in the ideal case, should now become \textit{isosceles}.

Actually, we experimentally observed a slight left-right imbalance of the output splitting ratio, in case of tritters fabricated with \textit{isosceles} triangle geometry. We also observed that the writing order of the three waveguides influenced this imbalance. This means that the inscription process of a second waveguide close to another one, already fabricated, may perturb slightly the first one or, alternatively, may be slightly disturbed by it. One can compensate this asymmetry by means of a small translation $\Delta$ [Supplementary Fig. S\ref{fig:crossSect} (b)] of the third waveguide.
For this reason, a better balance between the coupling coefficients can be obtained by adjusting the position of the third waveguide not only in the vertical direction [distance $h$ in Supplementary Fig. S\ref{fig:crossSect} (b)], but also in the horizontal direction (distance $\Delta$ in the same figure): the waveguides lay at the vertices of a \textit{scalene} triangle. 

To experimentally perform the optimization of the position of the third waveguide, we fabricated several tritters, with fixed $d$ and $L$, but different values of $h$ and $\Delta$. The output power repartition was measured for coherent light injection in the different input ports. This allowed to interpolate the values of $h$ and $\Delta$ which gave a balanced coupling.

A final adjustment was performed again on the interaction length $L$, both because of day-to-day reproducibility issue of the fabrication process, and because the presence of the third waveguide also may perturb the other two. The tritter chosen to be employed in the quantum experiments yields the following geometric parameters: $d = 11 \mu$m, $L = 2.9$~mm, $h = 10.3~\mu$m, $\Delta = 0.9~\mu$m.

Note that this optimization process required the fabrication of tens of tritters and couplers, with different geometric parameters. Given the fast processing speeds of femtosecond laser micromachining and since it doesn't require costly lithographic masks, the whole procedure resulted extremely cost-effective and required only a few days of work.

\section*{Supplementary Note 3: Reconstruction of the unitary matrix of the tritter}

Here we report the reconstruction of the tritter's transition matrix. The present device has been optimized to work at $\lambda = 795$ nm, and hence we performed its characterization at this wavelength. The experimental reconstruction of the transition matrix of the implemented tritter has been obtained by adopting a method based on the measurement of all possible combinations of two-photon interference terms [28]. 

Let $\mathcal{U}$ be the transfer matrix of the chip, whose elements $\mathcal{U}_{i,j}$ describe the transition amplitude of a photon entering in input port $i$ and exiting from output port $j$. In the classical case the probability to detect a photon out of each output $k$ and $l$, when two photons are injected in the inputs $i$ and $j$ where $i\neq j=1,2,3$ and $k \neq l=1,2,3$, is given by:
\begin{equation}
P^{\mathrm{C}}_{i,j;k,l}=\underbrace{\vert \mathcal U_{i,k} \mathcal U_{j,l}\vert^{2}}_{(a)}+\underbrace{\vert \mathcal U_{i,l} \mathcal U_{j,k}\vert^{2}}_{(b)}.
\end{equation}
The two terms (a) and (b) correspond respectively to:
\begin{enumerate}
\item[(a)] the photon entering in $i$ goes out of $l$ while the photon entering in $j$ goes out of $k$,
\item[(b)] the photon entering in $i$ goes out of $k$ while the photon entering in $j$ goes out of $l$.
\end{enumerate}
The above expression describes the classical case, since the overall probability is given by the sum of the probabilities for cases (a) and (b) and no interference occurs between the two paths. 

For two indistinguishable photons entering in the input ports $i$ and $j$, the probability to detect a photon in the output ports $k$ and $l$ is given by:
\begin{equation}
P^{\mathrm{Q}}_{i,j;k,l}=\vert \mathcal U_{i,k} \mathcal U_{j,l} + \mathcal U_{i,l} \mathcal U_{j,k}\vert^{2},
\end{equation}
corresponding to quantum interference between paths (a) and (b). By varying the relative time delay between the two impinging photons, we observe the Hong-Ou-Mandel effect leading to peaks (or dips) in the output coincidence events. The visibility $\mathcal V$ of the non-classical peak (or dip) is given by:
\begin{equation}
\mathcal V_{i,j;k,l}=\frac{P^{\mathrm{C}}_{i,j;k,l}-P^{\mathrm{Q}}_{i,j;k,l}}{P^{\mathrm{C}}_{i,j;k,l}},
\end{equation}
where positive values indicate a dip and negative values indicate a peak.

As shown in Ref. [28], the two-photon Hong-Ou-Mandel effect can be exploited to reconstruct the transition matrix of the tritter. We model the transition matrix as:
\begin{equation}
\mathcal{U} = \begin{pmatrix} \vert \mathcal{U}_{1,1} \vert && \vert \mathcal{U}_{1,2} \vert && \vert \mathcal{U}_{1,3} \vert \\ \vert \mathcal{U}_{2,1} \vert && \vert \mathcal{U}_{2,2} \vert e^{\imath \phi_{2,2}}&& \vert \mathcal{U}_{2,3} \vert e^{\imath \phi_{2,3}} \\ \vert \mathcal{U}_{3,1} \vert && \vert \mathcal{U}_{3,2} \vert e^{\imath \phi_{3,2}} && \vert \mathcal{U}_{3,3} \vert e^{\imath \phi_{3,3}} \end{pmatrix}.
\end{equation}
Here, only four free phases (from a global number of nine) are allowed to be different from zero. Indeed, a global phase can be neglected and other four phases can be set to zero by multiplying rows and columns for a phase factor, since the phase of the input state is undefined and photon-counting measurements are performed. 
The reconstruction then proceeds in two steps.

\textbf{\textsl{Step (1)}}. - The squared moduli of the matrix elements $\vert \mathcal U_{i,j} \vert^{2}$ are measured by feeding the tritter with single photons in only one input port. For each input $i$ we measured the count rate $n_{i,j}$ in each output port $j$. The squared moduli of the transition matrix elements can be calculated as:
\begin{equation}
\vert \mathcal U_{i,j}\vert^{2} = \frac{n_{i,j}}{\sum_{j=1}^{3}n_{i,j}}.
\end{equation}
The measured couplings are:
\begin{equation}
\label{eq:matrixmis}
\vert \mathcal U_{i,j}\vert^{2} =
\begin{pmatrix}
0.368 \pm 0.002 & 0.339 \pm 0.002 & 0.293 \pm 0.002 \\
0.317 \pm 0.002 & 0.310 \pm 0.002 & 0.373 \pm 0.002 \\
0.282 \pm 0.002 & 0.378 \pm 0.002 & 0.340 \pm 0.002
\end{pmatrix}
\end{equation}
The phases of the starting matrix are chosen to be equal to the theoretical values reported in Supplementary Eq. (\ref{eq:matrixteo}).

\textbf{\textsl{Step (2)}}. - In order to fully reconstruct the matrix $\mathcal{U}$, we measure the two-photon Hong-Ou-Mandel visibilities $\mathcal V^{m}_{i,j;k,l}$ corresponding to all possible combinations of two-fold correlations for all input states ($\vert 1_{1},1_{2},0_{3}\rangle$, $\vert 1_{1},0_{2},1_{3}\rangle$ and $\vert 0_{1},1_{2},1_{3}\rangle$, where $1,2,3$ are the input ports) [see Supplementary Fig. S\ref{fig:visibilities_U} (a)]. Then, we minimize numerically the distance between the theoretical visibility matrix $\mathcal V^{r}_{i,j;k,l}$, calculated as a function of the transition matrix elements, and the measured visibility matrix $\mathcal V^{m}_{i,j;k,l}$:
\begin{equation}
\label{eq:RMS}
\mathrm{RMS}=\sum_{i\neq j=1}^{3}\sum_{k\neq l=1}^{3}\frac{\left(\mathcal V^{r}_{i,j;k,l} - \frac{\mathcal V^{m}_{i,j;k,l}}{q}\right)^{2}}{\left(\sigma^{m}_{i,j;k,l}\right)^{2}}.
\end{equation}
Here $q$ is the factor which corresponds to the purity of the two-photon source, and $\sigma^{m}$ is the error matrix associated to the measured visibilities. The factor $p$ is evaluated by measuring the two-photon Hong-Ou-Mandel interference visibility in a symmetric 50-50 beam-splitter, which in our case corresponds to the value $q=0.94$.  The starting point $\mathcal{U}^{r}_{0}$ for the matrix $\mathcal U^{r}$ is given by adopting the measured splitting ratios of Supplementary Eq. (\ref{eq:matrixmis}) and by setting the phases $e^{\imath \phi_{i,j}}$ to their theoretical values, leading to:
\begin{equation}
\mathcal U_{0}^{r}= \begin{pmatrix}
0.608 & 0.583 & 0.538\\
0.566 & 0.557 e^{\imath 2 \pi/3} & 0.608 e^{\imath 4 \pi/3}\\
0.529 & 0.616 e^{\imath  4 \pi/3}& 0.583 e^{\imath  8 \pi/3}
\end{pmatrix}.
\end{equation}
The reconstructed matrix is shown in the main text and takes the following form:
\begin{equation}
\label{eq:matrixuni}
\mathcal U^{r}= \begin{pmatrix} 
0.593 \pm 0.001 & 0.5928 \pm 0.0008 & 0.5444 \pm 0.0006\\
0.5489 \pm 0.0005 & (0.5811 \pm 0.0006) e^{\imath (2.123 \pm 0.002)}& (0.6008 \pm 0.0006) e^{- \imath (2.030 \pm 0.002)} \\
0.5886 \pm 0.0009 & (0.5575 \pm 0.0007) e^{-\imath (2.167 \pm 0.002)} & (0.5853 \pm 0.0006) e^{\imath (2.110 \pm 0.002)}\end{pmatrix}.
\end{equation}
The errors on the parameters of $\mathcal{U}^{r}$ has been evaluted by performing $M=40$ times a numerical simulation of the reconstruction process with random gaussian noise on the input visibilities $\mathcal{V}^{m}_{i,j;k,l}$, of variance equal to the experimental error $\sigma^{m}_{i,j;k,l}$. As a figure of merit to quantify the quality of the algorithm we adopted the similarity  $\mathcal{S}_{r,m}=1-\sum_{i\neq j} \sum_{k \neq l} \vert q \mathcal{V}^{r}_{i,j;k,l} - \mathcal{V}^{m}_{i,j;k,l}\vert/(2 n)$ between the measured visibility matrix $\mathcal{V}^{m}$ [Supplementary Fig. S\ref{fig:visibilities_U} (a)] and the visibility matrix $\mathcal{V}^{r}$ [Supplementary Fig. S\ref{fig:visibilities_U} (b)]. We obtained a value $\mathcal{S}^{r,m}=0.989 \pm 0.002$. The correspondence between the reconstructed matrix $\mathcal{U}^{r}$ and the ideal matrix $\mathcal{U}^{t}$ of Supplementary Eq. (\ref{eq:matrixteo}) is evaluated from the similarity $\mathcal{S}_{r,t} = 1-\sum_{i\neq j} \sum_{k \neq l} \vert \mathcal{V}^{r}_{i,j;k,l} - \mathcal{V}^{t}_{i,j;k,l}\vert/(2 n)$, leading to: $\mathcal{S}_{r,t}=0.9768 \pm 0.0004$. These results show that the reconstructed matrix is in high agreement with the theoretical one. Finally, we repeated the reconstruction process by assuming a pure two-photon source ($q=1$). The similarities in this case are respectively $\tilde{\mathcal{S}}_{r,m} = 0.974 \pm 0.002$ and $\tilde{\mathcal{S}}_{r,t} = 0.9767 \pm 0.0004$, showing that the correction for the purity of the two-photon source is only a minor correction.

\subsection*{A. Reconstruction for input wavelength $\lambda=785$ nm}

The three-photon interference experiment described in the main text was performed by sending input states at $785$ nm. In order to verify the quality of the tritter at this wavelength, we performed the characterization of the device at this wavelength by adopting the same method described in the previous section. 
The results of the reconstruction are shown in Supplementary Figs. S\ref{fig:visibilities_U_785} and S\ref{fig:reconstruction_785}, leading to the following reconstructed matrix:
\begin{equation}
\label{eq:matrixuni785}
\mathcal{U}^{' \, r}= \begin{pmatrix} 
0.656 \pm 0.001& 0.5439 \pm 0.0006 & 0.5233 \pm 0.0006 \\
0.5302 \pm 0.0006 & (0.6135 \pm 0.0006) e^{\imath (2.210 \pm 0.002)}& (0.5852 \pm 0.0004) e^{- \imath (2.077 \pm 0.001)}\\
0.5371 \pm 0.0007 & (0.5725 \pm 0.0003) e^{- \imath (2.128 \pm 0.001)}& (0.6194 \pm 0.0008) e^{\imath (2.188 \pm 0.001)}.
\end{pmatrix}
\end{equation}%
We then evaluated the similarity $\mathcal{S'}_{r,m}$ between the measured visibility matrix $\mathcal{V}^{' \, m}$ [Supplementary Fig. S\ref{fig:visibilities_U_785} (a)] and the reconstructed visibility matrix $\mathcal{V}^{' \, r}$ [Supplementary Fig. S\ref{fig:visibilities_U_785} (b)], leading to a value $\mathcal{S'}_{r,m}=0.976 \pm 0.001$. The correspondence between the reconstructed matrix $\mathcal{U}^{' \,r}$ and the ideal matrix $\mathcal{U}^{t}$ of Supplementary Eq. (\ref{eq:matrixteo}) is evaluated from the similarity $\mathcal{S'}_{r,t}$, leading to $\mathcal{S}_{r,t} = 0.9595 \pm 0.0003$. Hence, the reconstructed matrix keeps a high agreeement with an ideal tritter also for $785$ nm input photons. Analogously to the characterization at $795$ nm, the reconstruction process by assuming a pure two-photon source ($q=1$) leads to negligible modifications to the similarities: $\tilde{\mathcal{S}}^{\prime}_{r,m} = 0.972 \pm 0.001$ and $\tilde{\mathcal{S}}^{\prime}_{r,t} = 0.9588 \pm 0.0003$.

\section*{Supplementary Note 4: Modeling the experiment}

In this section we discuss a theoretical model that takes into account photon distinguishability and six photon terms in the theoretical prediction of the three-photon interference visibilities. Furthermore, we explicitly evalute the classical bounds for the Hong-Ou-Mandel visibilities by considering three independent, phase-randomized, input coherent states in the tritter.

\subsection*{A. Photon distinguishability and six-photon states}

Due to the spectral properties of parametric down-conversion sources, photons belonging to different photon pairs are not completely indistinguishable. More specifically, parametric down-conversion with broadband pump leads to the presence of spectral correlations between the generated photons, which reduce the purity of the generated photons. This limits the interference visibility for two photons belonging to different photon-pairs. The adoption of narrow interferential filters can help in reducing the amount of spectral correlations, and thus increasing the purity, at the cost of a decrease in detected signal. In our case, we adopted $\Delta \lambda = 3$ nm filters for a pump beam of bandwidth $\Delta \lambda_{p}=2$ nm. The overall spectral effect can be directly measured by performing an Hong-Ou-Mandel interference measurement in a 50/50 beam-splitter with two photons belonging to different photon pairs, leading to $V^{(2)}=0.63 \pm 0.03$.

To include spectral correlations in the theoretical predictions of the experimental visibilities, we considered as input state in the tritter the following density matrix:
\begin{equation}
\varrho = r \vert 1,1,1 \rangle \langle 1,1,1 \vert + (1-r) \vert 1_{a},1_{b},1_{a} \rangle \langle 1_{a},1_{b},1_{a} \vert.
\end{equation}
Here, $r$ is the parameter which takes into account the distinguishability between the three photons, and the indexes $i=a,b$ label the photon on input port $2$ belongs to a different photon pair with respect to the photons on input ports $1$ and $3$. Note that, differently from experiments involving only two-photon interference [39-41], the parameter $r$ is proportional to the overlap between the spectral functions of the three photons, and reads $r=p^2$, where $p$ is the indistinguishability of two photons belonging to different photon pairs. A direct estimate of $p$ is given by the measured visibility $V^{(2)}$.

Furthermore, parametric down-conversion sources in this regime present a non-negligible probability of generating three photon pairs. The overall state produced up to third order in the parametric gain $g$ is given by:
\begin{equation}
\vert \psi \rangle \simeq \vert 0 \rangle + \sqrt{2}\, g \vert \psi_{2} \rangle + \sqrt{3} \, g^{2} \vert \psi_{4} \rangle + 2 g^{3} \vert \psi_{6} \rangle,
\end{equation}
where $\vert \psi_{2} \rangle$, $\vert \psi_{4} \rangle$, and $\vert \psi_{6} \rangle$ correspond respectively to the two-photon, four-photon and six-photon contributions. In the present implementation, $g$ reaches a value of $g \simeq 0.12$. We then included in our theoretical prediction for the Hong-Ou-Mandel visibilities the contribution given by the six-photon term $\vert \psi_{6} \rangle$.

The effect of these experimental imperfections on the output probabilities is shown in Supplementary Fig. S\ref{fig:prob}. We compare the results for the expected outcome probabilities for a three-photon input state $\vert 1,1,1 \rangle$ in different cases: (i) an ideal tritter $\mathcal{U}^{t}$ with perfectly indistinguishable photons ($p=1$), (ii) the implemented tritter $\mathcal{U}^{r}$ with perfectly indistinguishable photons ($p=1$), (iii) an ideal tritter $\mathcal{U}^{t}$ with photon distinguishability ($p$), (iv) the implemented tritter $\mathcal{U}^{r}$ with photon distinguishability ($p$), (v) the implemented tritter $\mathcal{U}^{r}$ with photon distinguishability ($p$) and six photon contributions. We observe that the predictions for an ideal and the implemented one closely resemble each other showing the quality of the fabrication process. The main source of deviations from the ideal three-photon interference is given by the photon distinguishability, which partially modifies the $P^{\mathrm{q}}_{i,j,k}$ and $P^{\mathrm{(2)}}_{i,j,k}$ (B) distributions. Finally, the six photon contribution is shown to be only a small correction. 

\subsection*{B. Classical bounds for the Hong-Ou-Mandel visibilities}

We are now interested in studying the behaviour of the experimentally realized tritter when it is injected by classical light (i.e. independent phase randomized coherent states), in order to see whether or not the three-photon visibilities obtained by injecting a three single photons state can be ascribed to a quantum behaviour.  To this aim we start writing the input-output relations as a function of the reconstructed transfer matrix $\mathcal{U}^{r}$:
\begin{equation}
b^{\dag}_{i}=\sum_{j}\mathcal{U}^{\prime \, r}_{ij}a^{\dag}_{j},
\end{equation}
\noindent with $i=1,2,3$, and consider the evolution of the input state $\ket{\alpha_{1}}\ket{\alpha_{2}}\ket{\alpha_{3}}=D(\alpha_{1})D(\alpha_{2})D(\alpha_{3})\ket{0}$, where $\ket{\alpha_{i}}$ is a generic coherent state and $D(\alpha_{i})$ the corresponding displacement operator. By definition: $D(\alpha_{i})=e^{\alpha_{i}a^{\dag}_{i}-\alpha^{*}_{i}a_{i}}$, where $\alpha_{i}=|\alpha_{i}|e^{i\theta_{i}}$. By expanding the $a^{\dag}_{i}$ operators as a function of the $b_{i}$ ones, we can write the output state as:
\begin{equation}
\ket{\psi^{\mathrm{out}}}=\prod_{i=1}^{3}\sum_{n_{i}}e^{-|\gamma_{i}|^2/2}\frac{\gamma_{i}^{n_{i}}}{n_{i}!} b_{i}^{\dag n_{i}}\ket{0},
\end{equation}
\noindent where $\gamma_{i}=\sum_{j}[(\mathcal{U}^{\prime \, r})^{-1}]_{ji}\alpha_{j}$. The output probabilities related to the $(300,111,210)$ output configurations can then be obtained by projecting $\ket{\psi^{\mathrm{out}}}$ on the respective output state, we then obtain:
\begin{eqnarray}
P^{(3)}(111) & = &  e^{-\sum_{i=1}^{3}|\gamma_{i}|^{2}}\prod_{i=1}^{3}|\gamma_{i}|^{2}, \\
P^{(3)}(210) & = & \frac{1}{2!} e^{-\sum_{i=1}^{3}|\gamma_{i}|^{2}}|\gamma_{1}|^{4}|\gamma_{2}|^{2},\\
P^{(3)}(300) & = & \frac{1}{3!} e^{-\sum_{i=1}^{3}|\gamma_{i}|^{2}}|\gamma_{1}|^{6}.
\end{eqnarray}
\noindent these values correspond to the three photons perfect interference, being the three coherent states simultaneous inside the tritter. In order to find the corresponding dip (or peak) visibilities, we have to quantify the values of probabilities related to the delayed case in which one (or two) of the three coherent state is (are) delayed. If only one input state is delayed we obtain:
\begin{equation}
P^{(2)}(m_{1},m_{2},m_{3})=\frac{e^{-\sum_{i=1}^{3}|\gamma_{i}|^{2}}}{m_{1}!m_{2}!m_{3}!}\prod_{i=1}^{3}\left(|\gamma_{i}^{\prime}|^{2}+|\gamma_{i}^{\prime \prime}|^{2}\right)^{m_{i}},
\end{equation}
\noindent where
\begin{itemize}
\item for delayed state in port 2: $\gamma_{i}^{\prime}=\sum_{j=1,j \neq2}^{3}[(\mathcal{U}^{\prime \, r})^{-1}]_{ji}\alpha_{j}$ and $\gamma_{i}^{\prime \prime}=[(\mathcal{U}^{\prime \, r})^{-1}]_{2i}\alpha_{2}$,
\item for delayed state in port 3: $\gamma_{i}^{\prime}=\sum_{j=1,j\neq3}^{3}[(\mathcal{U}^{\prime \, r})^{-1}]_{ji}\alpha_{j}$ and $\gamma_{i}^{\prime \prime}=(\mathcal{U}^{\prime \, r})^{-1}]_{3i}\alpha_{3}$.
\end{itemize}
\noindent  When two over three coherent states are delayed:
\begin{equation}
P^{(1)}(m_{1},m_{2},m_{3})=\frac{e^{-\sum_{i=1}^{3}|\gamma_{i}|^{2}}}{m_{1}!m_{2}!m_{3}!}\prod_{i=1}^{3}\left(|\gamma_{i}^{\prime}|^{2}+|\gamma_{i}^{\prime \prime}|^{2}+|\gamma_{i}^{\prime \prime \prime}|^{2}\right)^{m_{i}},
\end{equation}
\noindent where $\gamma_{i}^{\prime}=[(\mathcal{U}^{\prime \, r})^{-1}]_{1i}\alpha_{1}$, $\gamma_{i}^{\prime \prime}=[(\mathcal{U}^{\prime \, r})^{-1}]_{2i}\alpha_{2}$,$\gamma_{i}^{\prime \prime \prime}=[(\mathcal{U}^{\prime \, r})^{-1}]_{3i}\alpha_{3}$.
\noindent Finally we perform an average over the three phases $\theta_{i},\; i=1,2,3$, since we are dealing with phase randomized coherent states:
\begin{equation}
\Gamma^{(k)}(m_{1},m_{2},m_{3})=\frac{1}{(2\pi) ^{3}}\int_{0}^{2\pi}\int_{0}^{2\pi}\int_{0}^{2\pi} P^{(k)}(m_{1},m_{2},m_{3}) d\theta_{1}d\theta_{2}d\theta_{3}.
\end{equation}
From these $\Gamma^{(k)}$ values, we can obtain the classical visibilities:
\begin{equation}
V=\frac{\Gamma_{\infty}-\Gamma_{0}}{\Gamma_{\infty}},
\end{equation}
\noindent where:
\begin{itemize}
\item when one port  is delayed: $\Gamma_{\infty}=\Gamma^{(2)}(m_{1},m_{2},m_{3})$ and $\Gamma_{0}=\Gamma^{(3)}(m_{1},m_{2},m_{3})$
\item when two ports are delayed: $\Gamma_{\infty}=\Gamma^{(1)}(m_{1},m_{2},m_{3})$ and $\Gamma_{0}=\Gamma^{(2)}(m_{1},m_{2},m_{3})$.
\end{itemize}

\section*{Supplementary references}

\begin{small}

\noindent [36] Estes, L., Keil, T. \& Narducci, L. 
Quantum-mechanical description of two coupled harmonic oscillators.
\emph{Phys. Rev.} \textbf{175}, 286-299 (1968).

\noindent [37] Szameit, A., Dreisow, F., Pertsch, T., Nolte, S. \&
Tuennermann, A. Control of directional evanescent coupling in fs 
laser written waveguides. \emph{Opt. Express} \textbf{15}, 1579-1587 (2007).

\noindent [38] Yariv, A. Coupled-mode theory for guided-wave optics. \emph{IEEE J. Quant. Electron.}
 \textbf{9}, 919-933 (1973).

\noindent [39] Tsujino, K., Hofmann, H.~F., Takeuchi, S. \& Sasaki, K. Distinguishing genuine entangled two-photon-polarization states 
from independently generated pairs of entangled photons. \emph{Phys. Rev. Lett.} \textbf{92}, 153602 (2004).

\noindent [40] Lu, C.-Y. \emph{et~al.} Experimental entanglement of six photons in graph states.
\emph{Nat. Phys.} \textbf{3}, 91-95 (2007).

\noindent [41] Yao, X.-C. \emph{et~al.} Observation of eight-photon entanglement.
\emph{Nat. Photon.} \textbf{6}, 225-228 (2012).

\end{small}